\documentclass[11pt]{article}
\pdfoutput=1
\usepackage{graphicx,amssymb,amsmath,amsfonts}
\usepackage{hyperref}
\usepackage[margin=1.2in]{geometry}
\usepackage[noadjust]{cite}
\usepackage[utf8]{inputenc}

\def\be{\begin{equation}}
\def\ee{\end{equation}}

\numberwithin{equation}{section}
\def\bea{\begin{eqnarray}}
\def\eea{\end{eqnarray}}

\newcommand{\alp}{\ensuremath{\alpha^\prime}}

\newcommand{\ccb}{c\bar{c}}

\newcommand{\MEV}{\text{ MeV}}
\newcommand{\GEVm}{\text{ GeV}^{-2}}
\newcommand{\GEV}{GeV\(^{-2}\)}

\newcommand{\BV}{{\it{V-baryonium}}\,}

\begin{document}
\begin{titlepage}
\title{\textbf{Deciphering the recently discovered tetraquark candidates around 6.9 GEV}}

\author{\textbf{Jacob Sonnenschein}\(^{[a]}\) \\ \href{mailto:cobi@tauex.tau.ac.il}{cobi@post.tau.ac.il} \and \textbf{Dorin Weissman}\(^{[b]}\) \\ \href{mailto:dorin.weissman@oist.jp}{dorin.weissman@oist.jp}}

\date{\(^{[a]}\)\emph{The Raymond and Beverly Sackler School of Physics and Astronomy},\\
	\emph{Tel Aviv University, Ramat Aviv 69978, Tel Aviv, Israel} \\[1.3\baselineskip] \(^{[b]}\)\emph{Okinawa Institute of Science and Technology},\\
	\emph{1919-1 Tancha, Onna-son, Okinawa 904-0495, Japan}  \\[1.5\baselineskip] \today}
	
\maketitle
\begin{abstract}
Recently a novel hadronic state of mass 6.9 GeV, that decays mainly to a pair of charmonia, was observed in LHCb.
The data also reveals a broader structure centered around 6490 MeV and suggests another unconfirmed resonance centered at around 7240 MeV, very near to the threshold of two doubly charmed \(\Xi_{cc}\) baryons. We argue in this note that these exotic hadrons are genuine tetraquarks and not molecules of charmonia. It is conjectured that they are \BV, namely,  have an inner structure of a baryonic vertex with a \(cc\) diquark attached  to it, which is connected by a string to an anti-baryonic vertex with a \(\bar c \bar c\) anti-diquark. We examine these states as the analogs of the \BV  states $\Psi(4360)$ and $Y(4630)$/$\Psi(4660)$ which are charmonium-like tetraquarks. One way to test these claims is by searching for a significant decay of the state at 7.2 GeV into $\Xi_{cc}\overline\Xi_{cc}$. Such a decay would be  the analog of the decay of the state $Y(4630)$ into to $\Lambda_c\overline\Lambda_c$. We further argue that there should be trajectories of both orbital and radial excited states of the $X(6900)$. We predict their masses. It is possible that a few of these states have already been seen by LHCb.
\end{abstract}

\end{titlepage}

\flushbottom
\section{Introduction}
\begin{figure} \centering \label{fig:peaks}
\includegraphics[width=0.60\textwidth]{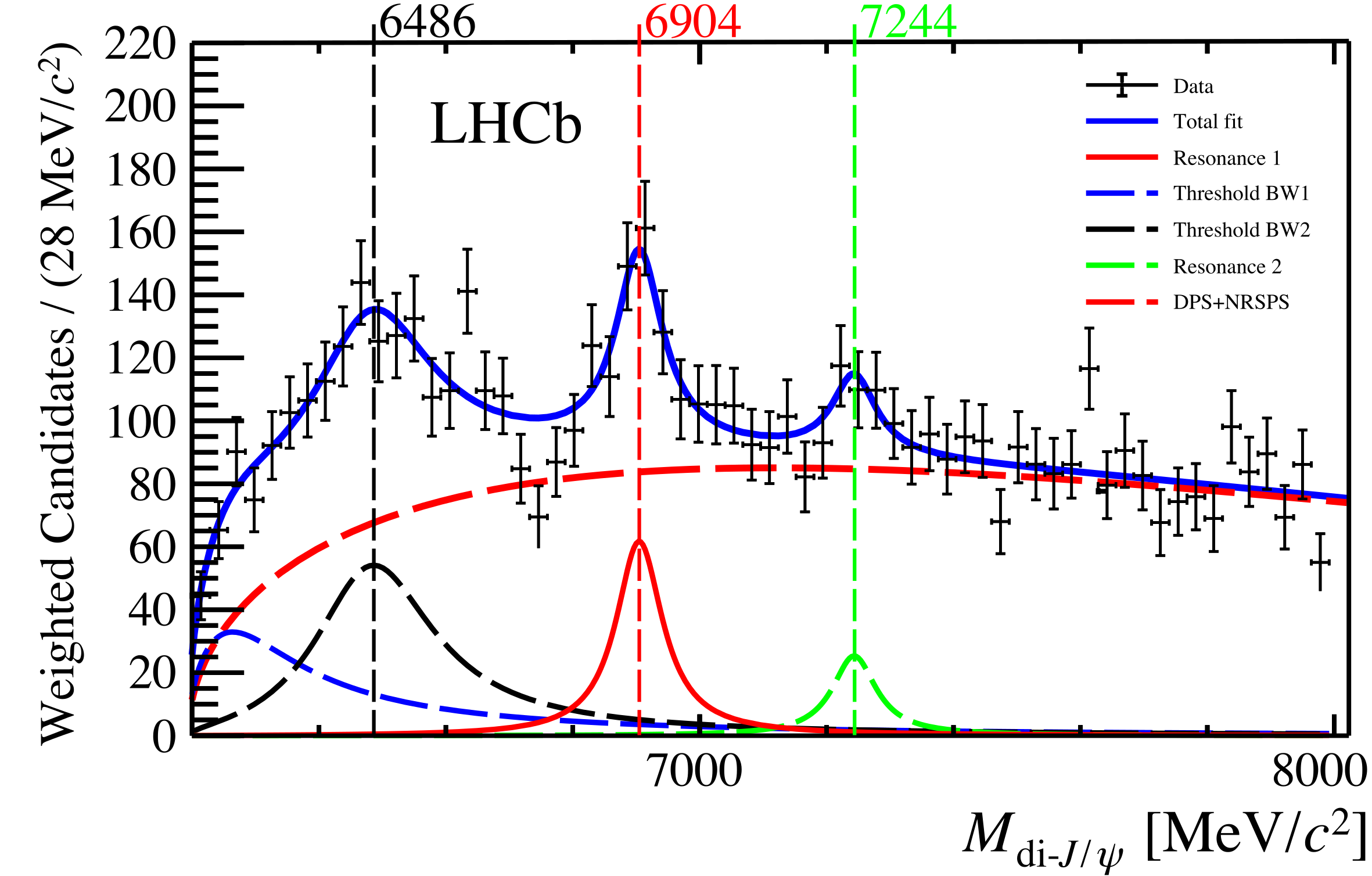}
\caption{Location of peaks in the LHCb data. Adapted from figure 7 in \cite{Aaij:2020fnh}. The 7.2 GeV state appears to be almost exactly on the \(\Xi_{cc}\bar \Xi_{cc}\) threshold, which is at 7242 MeV.}
\end{figure}

Recently, LHCb reported the discovery of a new structure in the \(J/\Psi\) pair mass spectrum at around 6.9 GeV \cite{Aaij:2020fnh}. This is a new exotic candidate, \(X(6900)\), which is expected to be a fully charm tetraquark, that is have the quark content \(cc\bar c \bar c\). Its measured mass and width are (in MeV)
\be M[X(6900)] = 6905\pm 11\pm 7 \ee
\[ \Gamma[X(6900)] = 80\pm 19\pm 33 \]
or, using a second fitting model
\be M[X(6900)] = 6886\pm 11\pm 11 \ee
\[ \Gamma[X(6900)] = 168\pm 33\pm 69\]
The data also reveals a broader structure centered around 6490 MeV, referred to as a ``threshold enhancement'' in \cite{Aaij:2020fnh}, and also suggests the existence of another resonance centered at around 7240 MeV, very near to the threshold of two doubly charmed \(\Xi_{cc}\) baryons. See figure \ref{fig:peaks}. The higher state, which we will refer to as \(X(7200)\), has not been confirmed at this stage.

In describing exotic hadrons one broadly distinguishes between ``molecules'' which are bound states of color singlets, and genuine exotic hadrons which cannot be described in that way. In  the case of the latter there should be a mechanism of building up  the exotic  hadron from its constituents. 

In \cite{Sonnenschein:2018fph} it was shown that the spectra of mesons and baryons of both light and heavy quark content match very nicely the spectra of the HISH (Holography inspired stringy hadron) model \cite{Sonnenschein:2016pim}.\footnote{In holography hadrons are described by stringy configurations in curved ten dimensional curved background. The HISH model is based on a map between  these strings and strings in flat four dimensional space, which are used to describe mesons, baryons, glueballs and exotics. For a holographic model of tetraquarks in a different approach see \cite{Liu:2019mxw,Liu:2019yye}.} 
%%%%%%%%%%%%%%%%%
A meson in HISH is a string with two massive particles (quarks) on its ends.  A baryon of this model is built from a baryonic vertex (BV)\footnote{In holography the BV is a $D_p$-brane wrapped over an $p$-cycle which by conservation of charge must be connected to $N_c$ strings \cite{Witten:1998xy}.} 
which is connected to three quarks. For many states in the spectrum the preferred configuration is such that the BV and two of the quarks are close together and form a diquark, which is attached by a string to the third quark. The BV is the holographic realization of the string junction of QCD, which was proposed in \cite{Rossi:1977cy,Montanet:1980te}.

\begin{figure}[ht!] \centering
	\includegraphics[width=0.60\textwidth]{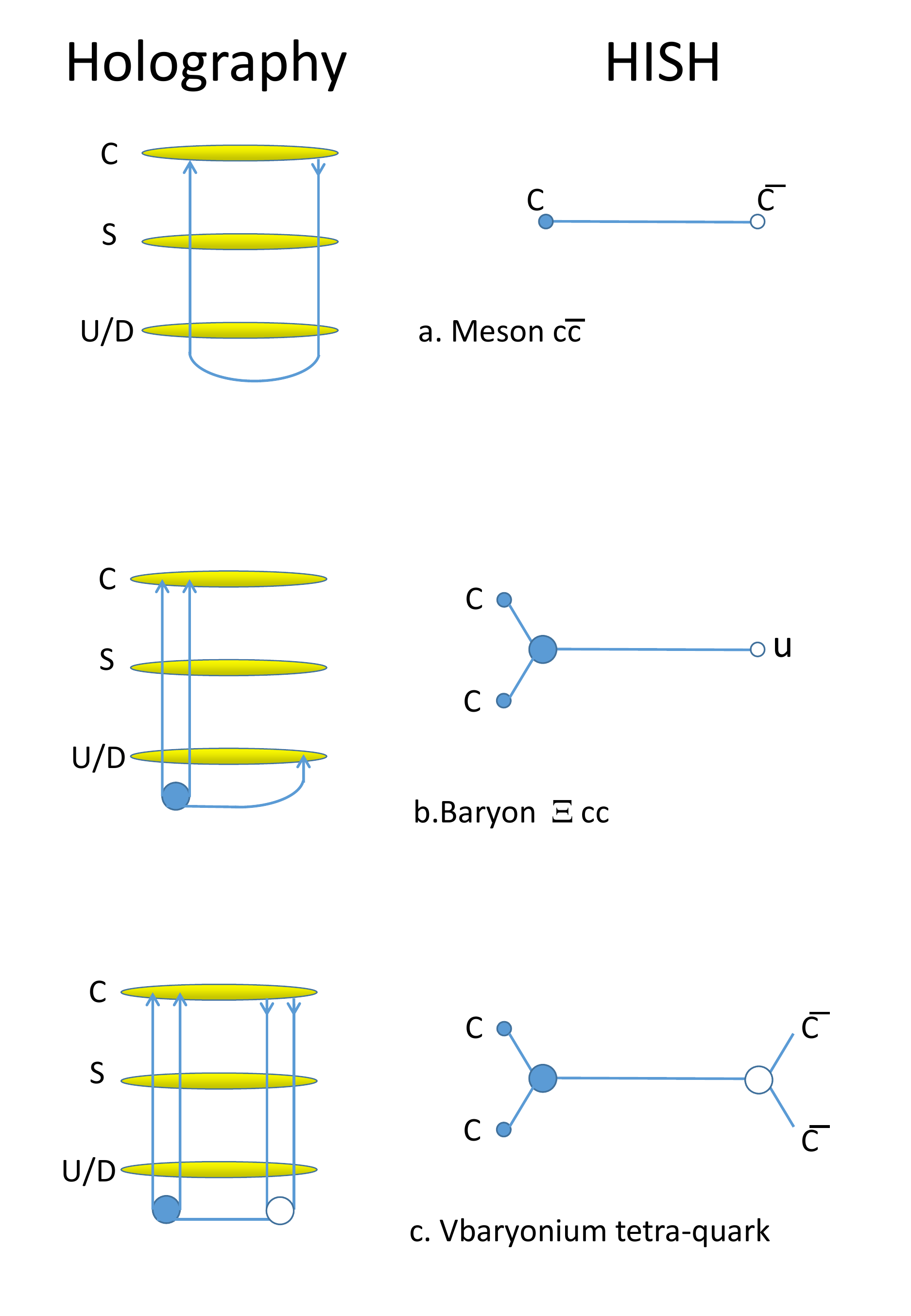}
			\caption{\label{fig:HISH} The structure of hadrons in holography and HISH: (a) \(c\bar c\) meson, e.g. \(J/\Psi\). (b) Doubly charmed baryon, \(\Xi_{cc}\). (c) Fully charm \BV tetraquark, conjectured structure of \(X(6900)\).}
\end{figure}

Since the strings have an orientation, it can be used to define anti-quarks and anti-BV branes. With these basic ingredients of baryonic vertices, diquarks and strings, it is easy to construct a tetraquark. The basic picture is of a diquark connected by a string to an anti-diquark. Holographically it involves a baryonic vertex and an anti-BV that are connected by a string. Then, since the vertices need to have in total three strings incoming or outgoing, they are each connected to two more strings, which form the diquark and anti-diquark. The meson, baryon, and tetraquark are depicted in figure \ref{fig:HISH}. We refer to this type of tetraquark as \BV  since it can be viewed as a bound state of a BV plus diquark and an anti-BV plus anti-diquark, and the BV and anti-BV can be thought of carrying baryon number \(+1\) and \(-1\) respectively. Note that it is not a baryonium in the sense of a bound state of color singlets.

The \BV tetraquarks are characterized by the following properties:

\begin{enumerate}

\item Since it is built from a string, then like all other stringy hadrons it must have trajectories of higher excited states, one of states with higher angular momentum, and one of higher radial excitation number. We refer to this trajectory as a HISH modified Regge trajectory (HMRT) \cite{Sonnenschein:2018fph}. Since the mesons, baryons, and \BV tetraquarks can all be seen as a single string with massive particles on it, as depicted in figure \ref{fig:HISH}, the function describing the trajectories is the same for the different types of hadrons.

\item The ground state, and possibly a few lower excited states of the \BV, has a mass which is lower than the sum of masses of the lightest baryon and anti-baryon pair with the same diquark and anti-diquark content as that of the tetraquark. On the other hand higher excited states on its trajectory will have a mass larger than that threshold mass.

\item Correspondingly, there are different decay modes for the below and above threshold excited states. The most probable mechanism of decay of the higher excited states is via a breaking of the string the connects the BV and anti-BV, creating a light \(q\bar q\) pair. Then the decay products of the \BV are a pair of a baryon and an antibaryon \cite{Sonnenschein:2016ibx} that preserve the diquark and anti-diquark. The probability for this type of decay is proportional to the total length of the string.

\item States with masses below the threshold cannot decay by a breaking apart of the string. Instead we propose here another mechanism where due to quantum fluctuations the BV and anti-BV reach the same point and annihilate. The strings attached to them are then reconnected to form two mesons with the same quark content as the original tetraquark. In QCD terms the quarks simply rearrange and separate into two mesons. The probability of the vertices meeting is suppressed exponentially in the string length squared.
\end{enumerate}

Our prototype case was the exotic hadron $Y(4630)$ discovered in Belle \cite{Pakhlova:2008vn} and seen to decay to \(\Lambda_c\overline\Lambda_c\). In \cite{Sonnenschein:2016ibx} we argued that it is a charmonium-like tetraquark built from a diquark of \(c\) and \(u/d\), and an anti-diquark of \(\bar c\) and \(\bar u/\bar d\). Naturally, one can built many other types tetraquarks from other diquarks and anti-diquarks. In \cite{Sonnenschein:2016ibx} we discussed the analogous bottomonium-like tetraquarks, as well as the possibility of a tetraquark containing \(s\bar s\).

In this note we revisit the charmonium-like \BV candidate, and apply the lessons learned from it is to the system of the newly discovered state \(X(6900)\) and the other resonances in its vicinity. We can use the string model to predict the masses and  evaluate the widths of the states.

In particular we show that there is good reason to believe that the state \(Y(4630)\) is the first radial excitation of \(\Psi(4360)\), which was seen in the \(\Psi(2S)\pi^+\pi^-\) channel. Another resonance seen in that channel is \(\Psi(4660)\), which we identify with the \(Y(4630)\) observed in \(\Lambda_c\overline\Lambda_c\). The \(Y(4630)\) is above the baryon-antibaryon threshold by about 60 MeV and has both types of decays described above. The two states are both \BV tetraquarks and belong on a HMRT together.

Similarly, we see that one can organize the three resonances at 6490, 6900, and 7240 MeV of figure \ref{fig:peaks} on a HMRT, and we can estimate their widths based on the stringy model.

We also conjecture the existence of many more \BV states grouped into symmetric, semi-symmetric and asymmetric tetraquarks, which should be located around the corresponding baryon-antibaryon thresholds. We further argue that finding certain such  tetraquarks can be an indirect way to observe for the first time baryons like those with bottom and charm quarks or doubly bottom baryons.

The note is organized as follows: In section \ref{sec:tetra} we briefly describe the structure and properties of the \BV tetraquarks in the HISH model. In section \ref{sec:classes} we describe the classes of tetraquarks.  In section \ref{sec:decays} we describe the two possible decay mechanisms for the \BV states. Section \ref{sec:cucu} is devoted to the analysis of the system of $\Psi( 4360)$/$Y(4630)$, the charmonium-like tetraquark candidates. In section \ref{sec:cccc} we present our analysis and conjectures regarding the recently discovered state $X(6900)$ and the other states in its vicinity. The structure and masses are described in section \ref{sec:ccccmasses}, and decay widths in \ref{sec:ccccdecay}. In section \ref{sec:novelbv} we present some predictions about novel \BV tetraquarks.  In \ref{sec:summary} we summarize our predictions and list several open questions. 
%%%%%%%%%%%%%%%%%%%%%%%%%%%%%%%%%%%%%%%%%%%%%%%%%%%%%  
\section{The structure and properties  of the \texorpdfstring{\BV}{V-baryonium} tetraquarks}\label{sec:tetra}
%\section{tetraquark configurations} \label{sec:tetra}
The construction of the HISH tetraquarks is based on the observation that many baryons in nature, and especially the excited ones, have the structure of a diquark connected by a string to a quark \cite{Sonnenschein:2014bia}. In holography this is realized as a baryonic vertex (BV) connected to three strings representing quarks. Two strings are very short and together with the BV form a diquark, and the last string is a long one connecting to the remaining quark.\footnote{In holography, for a diquark that includes a heavy flavor quark, the string that connects it to the BV is a long string in the holographic direction but its projection to real space coordinates is small.} This system depicted in figure \ref{fig:HISH}b. With this picture of the baryon it is natural to construct an exotic tetraquark. One can simply replace the quark at the end of the string with an anti-diquark, as depicted in figure (\ref{fig:HISH}c). Thus, a \BV tetraquark is a string where on one end of it there is a BV plus a diquark and on the other end an anti-BV and an anti-diquark. The \BV has ``hidden baryon number'' - its baryon number is zero but it it is constructed from objects carrying baryon numbers \(+1\) and \(-1\).

In the HISH model we depict the hadron as a string with tension \(T\) and with massive endpoints \(m_1\) and \(m_2\). In \cite{Sonnenschein:2014jwa} classical solutions of rotating strings with massive endpoints were written down. The corresponding energy and angular momentum of this classical system are given by
\be M = \sum_{i=1,2}\left(\gamma_i m_i + T\ell_i\frac{\arcsin{\beta_i}}{\beta_i}\right)\,, \ee
\be J = \sum_{i=1,2}\left[\gamma_i m_i \beta_i \ell_i + \frac12T\ell_i^2\left(\arcsin\beta_i-\beta_i\sqrt{1-\beta_i^2}\right)\right]\,.\ee
where \(\beta_i\) is the velocity of the endpoint, \(\gamma_i = (1-\beta_i^2)^{-1/2}\), and \(\ell_i\) is the radius of rotation of the endpoint. The total length of the string is \(L=\ell_1+\ell_2\), and the solution has to obey the equations
\be \frac{T}{\gamma_i} = \frac{m_i \gamma_i\beta_i^2}{\ell_i}\ee
for \(i=1,2\), which are the force equations on the endpoint particles. The endpoint velocities \(\beta_i\) are related to each other from the condition that the angular velocity is the same for both endpoints, implying
\be \omega = \frac{\beta_1}{\ell_1} = \frac{\beta_2}{\ell_2}\,.\ee
These are the defining equations of a function \(J(M)\), which we call the classical HMRT (holography modified Regge trajectory). For the case of massless endpoints this reduces to the famous classical  Regge trajectory relation $J= \alp M^2$ with the Regge slope \(\alp = (2\pi T)^{-1}\). Note that we typically associate \(J\) with the orbital angular momentum of a state, the spin of the endpoints being added separately. We reserve the notation \(L\) for the string length.

By comparing the predictions of the HISH model (including quantum corrections) with the known spectra of mesons and baryons the optimal values of the ``string endpoint masses''\footnote{The string endpoint mass corresponds in holography to the action of the string along the vertical segment of the hadronic string. The mass is roughly the tension times the length of the string in the holographic direction.} $m_i$ for the quarks and for the diquarks of the various flavored quarks were determined \cite{Sonnenschein:2018fph}. We assume here, based on our previous analysis of the baryonic spectra, that the mass of a diquark is given at leading order by $m_{q_1 q_2} = m_{q_1}+m_{q_2}$, even though in the holographic picture this is far from obvious due to the BV being part of the diquark.

The previous expressions are classical. Quantizing the fluctuations of the string around the classical solution yields for the case of the massless endpoints the quantum Regge trajectory $J+n = \alp M^2 + a$  where $\omega_n=n$ is the eigenfrequency  of the $n$-th excited state and $a=\frac12\sum_n \omega_n$ is referred to as the {\it intercept}. In \cite{Sonnenschein:2018aqf} the quantization of the string with small massive endpoints was performed in the non-critical dimensions $d=4$  and the eigenfrequencies and intercept were determined as a function of the endpoint masses.

For phenomenological purposes we found \cite{Sonnenschein:2017ylo,Sonnenschein:2018fph} that we should introduce the quantum corrections by the replacement \(J \to J+n - a\), in analogous fashion to the massless string. Since we add \(n\) and not \(\omega_n\) (which is only known for fluctuations around strings obeying \(TL/m \gg 1\)), the price we pay is that we get different effective slopes for orbital and radial trajectories, \(\alp_J \neq \alp_n\). As for the intercept, in bosonic string theory \(a=1\), which implies a tachyonic ground state. To fit the experimental data one always needs a negative intercept. Since we define it in relation to the orbital angular momentum, we can write  $\tilde a = a-S< 0$, where $S$ is the spin (not the total angular momentum) of the system and \(a\) the intercept defined with respect to the trajectory of the total angular momentum as a function of the mass. The negative intercept implies that effectively there is a repulsive Casimir force that acts on the string endpoints, $F_C=\frac{-\tilde a}{L^2} $ where $L$ is the length of the string. Due to this force non-rotating strings, that classically would collapse to zero size because of the string tension, now have a finite length and are non-tachyonic.
%%%%%%%%%%%%%%%%%%%%%%%%%%%%%%%%%%%%%
 
When comparing the intercepts of mesons and baryons we find that they are different even when they are composed of the same type of quarks. This implies that the fluctuations of a system that includes a string and an  endpoint built from a BV and a diquark is different from that with  an ordinary quark as an endpoint. There are two possible reasons for this difference: (i) the quantum fluctuations of the BV, and/or (ii) a change of the boundary condition of the string which changes the  eigenmodes of the fluctuations. If we denote by $\tilde a_m$ and $\tilde a_b$ the intercepts of a meson and a baryon of the same flavor structure,  then the difference of the intercepts is $\Delta\tilde a =\tilde a_b-\tilde a_m $. If the cause of the difference is the quantum fluctuations of the BV then we anticipate that 
$ \tilde a_t\equiv \tilde a_m + 2 \Delta \tilde a = \tilde a_b + \Delta \tilde a\,. \label{eq:intercept_tetra}$
%But the intercept of the tetraquark will not be of this form  if   it is due to the change of the  boundary conditions of the string. In any case, since we do not know how to determine this difference from string theory (recall that the BV is a wrapped $D_p$ brane over a $c_p$ cycle), 
In this work we simply use experimental data to determine the intercepts when necessary.

The parity and charge conjugation parity of the \BV should depend on the string and its endpoints. For mesons the rules $P=(-1)^{L+1}$, $C= (-1)^{L+S}$, which are based only on the endpoint quarks, are very well known. Constructing tetraquarks as a bound state of a diquark and an anti-diquark, then unlike for mesons the endpoints will be bosons of spin $0$ or $1$. Thus, now parity and charge conjugation should be $P = (-1)^L$ and $C= (-1)^{L+S}$.

\subsection{Classes of tetraquarks } \label{sec:classes}
%%%%%%%%%%%%%%%%%%%%%%%%%%%%%%%%%%%  
As was discussed in \cite{Sonnenschein:2016ibx}   tetraquark configurations  can be classified according to symmetries in their flavor content. We group them into symmetric, semi-symmetric and asymmetric states as follows:   
\begin{itemize}
\item {\bf Symmetric tetraquarks} where the anti-diquark is built from the anti-quarks associated with those found in the diquark. For instance  the diquark $cu$, and the anti-diquark $\bar c \bar u$. These symmetric tetraquark configurations are flavorless and carry zero electric charge. Altogether there are 15 different symmetric tetraquarks for 15 unique types of diquarks composed of quarks of five flavors.

\item {\bf Semi-symmetric tetraquarks} in which there is one pair of quark and antiquark of the same flavor and one pair which includes a quark and an anti-quark of different flavors, for instance \((cu)(\bar c \bar s)\).
Thus the flavor content of these exotic hadrons is the same as of mesons that carry non-trivial flavor and they can carry a charge of $+1$, 0, or $+1$. We have 5 possibilities for a matched pair and 20 for the unmatched pair, so altogether there are 100 possible semi-symmetric tetraquarks. 
%For Examples are shown  in  figure \ref{asymtetra} the left figure may describe a semisymmetric tetraquark with a matched pair of \(u\) and \(\overline u\), with a \(b\) and a \(\overline c\) as the unmatched pair.

\item {\bf Asymmetric tetraquarks}, where both pairs are of different flavor.
%\subsection{Asymmetric tetraquarks}
 We could have any pair of quark forming the diquark and any two antiquarks forming the anti-diquarks thus altogether there are a priori $15\times 15=225$ possibilities of tetraquarks. Out of the 225 tetraquarks, we have 15 that are symmetric, 100 semi-symmetric, and thus 110 are asymmetric ones. The asymmetric tetraquarks can carry a charge of $-2$, $-1$ , 0, $+1$, or $+2$, with a charge of \(\pm2\) being an obviously exotic feature (for hadrons with baryon number zero). These can be manifestly exotic due to flavor content as well, e.g. \((cs)(\bar u\bar d)\) for which a candidate was recently observed \cite{Aaij:2020ypa,Aaij:2020hon,Karliner:2020vsi}.
 
\end{itemize}

			%%%%%%%%%%%%%%%%%%%%%%%%%%%%%%%%%%%%%%%%%%%%%%%%%%%%%%%%%%%%
\section{Decays of the \texorpdfstring{\BV}{V-baryonium} tetraquarks} \label{sec:decays}
The generic decay of a stringy hadron is by breaking up of the string into two strings. However, for mesons we know that there are also decay channels that involve an annihilation of the string endpoints, as is the case for OZI suppressed decays for the low lying charmonium and bottomonium mesons \cite{Sonnenschein:2017ylo}.  

The decay of the low lying tetraquark states cannot be via breaking of the string since they can generally be below the relevant threshold. Their decays involve an annihilation, not of a quark-antiquark pair, but rather of the BV with the anti-BV.

The higher excited states on the \BV HMRT can break apart, and above threshold those are the dominant decays. We describe both types of decays in the following.
 
\subsection{The decay mechanism of below threshold states} \label{sec:31}
 As explained in the introduction the natural decay mode of the \BV tetraquark is into a baryon and anti-baryon. However, this cannot take place for the ground state and possibly other low lying states because  their  mass is  less than that of the baryon anti-baryon pair. This is expected to be a generic property of the \BV since it can be viewed as a  type of baryon anti-baryon bound state, but  where one pair of a quark and anti-quark was annihilated, making it is a state of four constituents and not six. Then there is a certain binding energy which can ascribed to these states and therefore they cannot decay into a the baryon anti-baryon pair. However, the \BV states are not stable since they can decay via other channels. A possible decay mechanism involves the annihilation of the BV with the anti-BV.
 
\begin{figure}[ht!] \centering
	\includegraphics[width=0.46\textwidth]{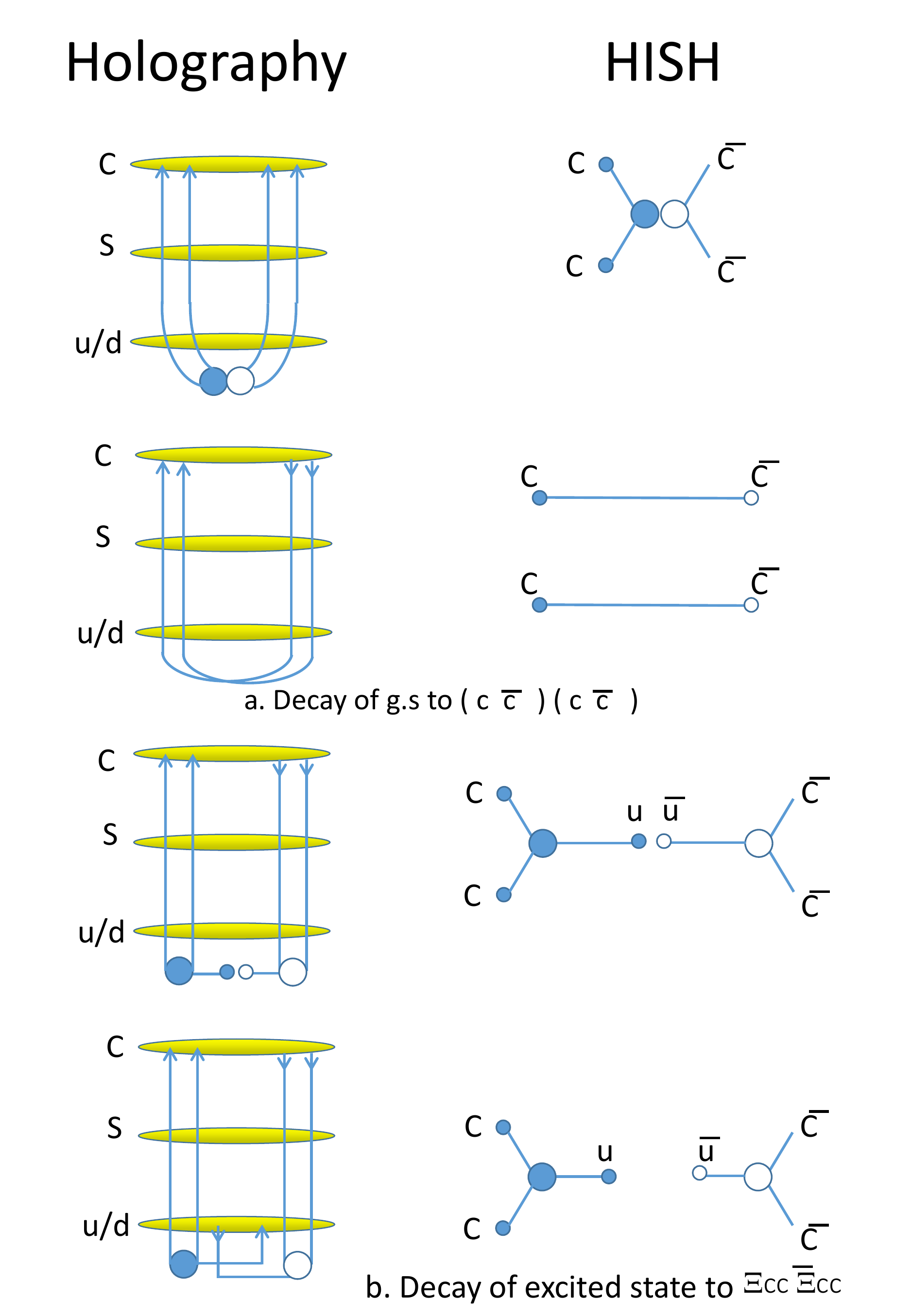}
				\caption{\label{fig:decaysf} (a) The decay of the ground state of a \(cc\bar c\bar c\) \BV tetraquark to a pair of charmonia. (b) The decay of an excited state to a doubly charmed baryon and anti-baryon, both in holography (left figures) and in HISH (right figures).}
			\end{figure}

Due to quantum fluctuations there is a certain probability that the BV and anti-BV, that on the average are separated at a  distance  $L$ (the length of the string), will hit each other. The annihilation process should be accompanied by a reconnection of the strings that originally connected the BV to the quarks on flavor branes with the strings that stretched from the anti-BV to the flavor branes but at another point in space. This mechanism of decay is depicted in figure \ref{fig:decaysf}a. In QCD terms the quarks have simply reorganized to form two mesons.

The probability of such a decay is the product of the probability that the BV  and anti-BV will be at the same point in space times the probability of an annihilation of the BV pair.
\be
{\cal P} = {\cal P}_{same\,location}\times {\cal P}_{annihilation}
\ee
In \cite{Sonnenschein:2017ylo} we estimated ${\cal P}_{same\,location}$ for a similar system to be 
\be
{\cal P}_{same\,location}=\sqrt{{\pi} {2T}}e^{\frac{-T L^2}{2}}\approx \sqrt{{\pi} {2T}} e^{-\frac{4(M-2m)^2}{9 T}} \label{eq:decaywidthannihilation}\ee
where $T$ is the string tension, $M$ is the mass of the tetraquark and $m$ is the mass of the BV and diquark.  The main result is the exponential suppression in the string length.

In the simplest process of reconnection of the strings following the annihilation, the decay products are two mesons with the same quark content as the original tetraquark. There are also processes where additional light \(q\bar q\) pairs are created and there are three or more mesons in the final state. An example is the decay of the charmonium-like \BV candidate \(\Psi(4360)\) into \(\Psi(2S)\pi^+\pi^-\) which we examine in section \ref{sec:cucu}. In that case we still expect the overall factor of \(\mathcal P_{\text{annihilation}} \sim e^{-TL^2/2}\) in the decay width.
%%%%%%%%%%%%%%%%%%%%%%%%%%%%%%%
\subsection{The decay mechanism of the higher excited states} \label{sec:32}
For \BV states with masses above the baryon-antibaryon threshold then the natural mode decay is via the breakup of the string. This is drawn in \ref{fig:decaysf}b. As was shown in \cite{Sonnenschein:2017ylo}, the total decay width of any hadron that involves a breaking up of a string is given by
\be\label{decaywidth}
\Gamma = \frac\pi2 A \times \Phi \times T L
\ee
Where $A$ is a dimensionless constant found to be $A\sim 0.1$ for mesons, $T$ and $L$ are the string tension and length. The factor \(\Phi\) accounts for phase space. As in the previous subsection, here also we can express the  string length $L$  in term of the mass of the tetraquarks and the masses of the string endpoint particles. The linearity of the decay width in $L$ can be intuitively understood since the string can be torn apart at any point along its length.

The phase space factor \(\Phi\) takes the form
\be \Phi = \frac{2|p_f|}{M} = \sqrt{\left(1-\left(\frac{M_1+M_2}{M}\right)^2\right)\left(1-\left(\frac{M_1-M_2}{M}\right)^2\right)} \ee
where \(M\) is the mass of the decaying particle, and \(M_1\) and \(M_2\) the masses of the outgoing particles, in the main channel of decay. It is only included in the phenomenological model to account for the suppression of decays for states that have little phase space to decay, so it is only relevant to states just above threshold to decay, and should not be included for states which have multiple viable decay channels (see section 8.1.2. in \cite{Sonnenschein:2017ylo} for a detailed explanation).

Lastly, we discuss the possibility that both types of decays discussed in this section are present. The simplest way to account for both is to add the two widths as
\be \Gamma = \frac\pi2 A\times \Phi \times TL + \tilde A \sqrt{T} \times e^{-\frac{TL^2}{2}} \label{eq:decaywidthboth}\ee
where \(\tilde A\) is another dimensionless constant, proportional to \(\mathcal P_{\text{annihiliation}}\) from the previous subsection. The former channel will be dominant for excited states, which are long strings, but could be suppressed by the phase space factor for near-threshold states.

%%%%%%%%%%%%%%%%%%%%%%%%%%%%%%%%%%%%%%%%%%%%%%%%%%%%
\section{The \texorpdfstring{$\Psi(4360)$}{Psi(4360)} and \texorpdfstring{$Y(4630)$}{Y(4630)} tetraquark system} \label{sec:cucu}
The sector of hidden charm mesons is particularly rich in exotic candidates \cite{Ali:2017jda,Yuan:2018inv,Brambilla:2019esw}. The prototype state that we believe is a \BV state is the \(Y(4630)\) which  was observed by Belle in 2008, in the process \(e^+e^-\to \gamma \Lambda^+_c\Lambda^-_c\), with a significance of \(8\sigma\) \cite{Pakhlova:2008vn}. The parameters measured there for the \(Y(4630)\) were
\be J^{PC} = 1^{--}\,,\,\, M_{Y(4630)} = 4634^{+9}_{-11}\,, \,\, \Gamma_{Y(4630)} = 92^{+41}_{-32} \,.\ee
In \cite{Sonnenschein:2016ibx} we considered several options of interpretation of this hadronic state and concluded that the most plausible possibility is that it is a \BV tetraquark. This conclusion is largely based on the fact that it decays to \(\Lambda_c\overline{\Lambda}_c\) which as mentioned above is the natural decay mode of such a state. The tetraquark nature of the \(Y(4630)\) state was explored in other contexts as well \cite{Cotugno:2009ys,Guo:2010tk,Brodsky:2014xia,Liu:2016sip,Guo:2016iej}.

The \(Y(4630)\) is often identified, including by the PDG \cite{PDG:2020}, with the nearby resonance \(\Psi(4660)\) (formerly \(Y(4660)\)), which was seen in \(e^+e^-\to\gamma\pi^+\pi^-\Psi(2S)\). It also has \(J^{PC}=1^{--}\), and its mass and width are, according to the latest average by the PDG \cite{PDG:Live},
\be M_{\Psi(4660)} = 4633\pm 7\,,\qquad \Gamma_{\Psi(4660)}=64\pm 9 \ee
Note that these averages also include the \(Y(4630)\) measurements of Belle.

We have two options. Either there are two separate states and the \(Y(4630)\) decays predominantly to baryon-antibaryon, or they are the same state that has both decays to \(\Psi(2S)\pi^+\pi^-\) and \(\Lambda_c\overline{\Lambda}_c\). In the following we expand upon the discussion in \cite{Sonnenschein:2016ibx} to include the latter possibility. Since the \(Y(4630)\) is only about 60 MeV above the $\Lambda_c\overline{\Lambda}_c$ threshold, it is more likely that it should have both channels of decay for its width to be as large as measured, as we show below. There have been multiple works studying the \(Y(4630)\) and \(\Psi(4660)\). These include \cite{Ebert:2008kb,Ding:2007rg,Wang:2016fhj,Anwar:2018sol,Sundu:2018toi,Yan:2018gik,Wang:2018rfw,Wang:2019iaa,Cao:2019wwt,Wang:2020prx,Ghalenovi:2020zen}.

Another role is played by the \(\Psi(4360)\), a lower mass state which has very similar properties. It is another \(1^{--}\) state seen in \(e^+e^-\to\gamma\pi^+\pi^-\Psi(2S)\). Its mass and width are measured to be
\be M_{\Psi(4360)} = 4368\pm 13 \MEV\,,\qquad \Gamma_{\Psi(4360)} = 96\pm 7 \MEV \ee
This state is well below the \(\Lambda_c\overline\Lambda_c\) threshold, but it is as wide or wider than the \(Y(4630)\).

The basic proposition of \cite{Sonnenschein:2016ibx} was that if the \(Y(4630)\) is a \BV tetraquark, it should be part of a HMRT. Using the values of \(m_{cu}\approx m_c\) and \(\alp\) of the \(\ccb\) trajectories, we used the known mass of the state \(Y(4630)\) to extrapolate from it to higher states along the trajectory. We use the values
\be m_c = 1490 \MEV\,,\qquad \alp\!_J = 0.86 \GEVm\,,\qquad \alp\!_n = 0.59 \GEVm\,. \ee
There are two different slopes, determined from the analysis of the charmonium spectrum: one for orbital trajectories in \(J\) and one for radial trajectories in \(n\).

By extrapolating the trajectory backwards to lower masses, we find that the \(\Psi(4360)\) is exactly at the right mass for the \(Y(4630)\) to be its first radially excited state.

We now suggest that \(Y(4630)\) and \(\Psi(4660)\) are the same state, which is a radially excited partner of the \(\Psi(4360)\), and the two states are both \BV tetraquarks. The \(\Psi(4360)\) is below threshold and decays via BV-anti-BV annihilation as described in section \ref{sec:31}, while the \(Y(4630)\) which is above but close to the \(\Lambda_c\overline\Lambda_c\) threshold has both types of decays.

For the higher excited states, the \(\Lambda_c\overline{\Lambda}_c\) decays should be dominant, and their masses should be such that they fall on the HMRT.

We can estimate the two partial decay widths of the \(Y(4630)\) in the following way. The decay via a breakup of the string is given by eq. \ref{decaywidth}. Writing it for \(Y(4630)\),
\be \Gamma_{\text{tear}} = \frac\pi 2 \times A \times \Phi(Y(4630)\to \Lambda_c\overline \Lambda_c) \times TL|_{Y(4630)} = A \times (268 \MEV) \lesssim 27 \MEV \ee
In the last step we use the typical value of \(A\) that we obtain from meson fits, which is around 0.1 or less.\footnote{There is an ambiguity in this calculation since we have two choices we can make for the tension \(T = (2\pi \alp)^{-1}\), corresponding to the two measured slopes. Since we take \(Y(4630)\) to be a radially excited state we use \(\alp_n\) in the calculation.}

This is not compatible with the full width of \(Y(4630)\) as measured by Belle. We can evaluate the partial width for \(\Psi\pi\pi\) based on the width of the \(\Psi(4360)\) as
\be \Gamma_{\text{annihilation}} = \Gamma[\Psi(4360)]\times \frac{\exp(-\frac{TL^2}{2})|_{Y(4630)}}{\exp(-\frac{TL^2}{2})|_{\Psi(4360)}} \approx 0.50\times \Gamma[\Psi(4360)] \approx 48 \MEV\ee
Now we can see that adding both channels as in eq. \ref{eq:decaywidthboth} gives a result that is quite close to the width of the \(\Psi(4660)\), which is measured at \(64\pm9\). By taking \(A\) to be \(\approx 0.06\) we get close to the exact measurement.\footnote{Fitting the same formulas with \(T=(2\pi \alp_J)^{-1}\) gives \(A=0.1\), which is more consistent with the other trajectories fitted in \cite{Sonnenschein:2017ylo}.}

In tables \ref{tab:pred_y_c_n} and \ref{tab:pred_y_c_j} we write the predicted masses of states on the radial and orbital HMRTs of the \(\Psi(4360)\) and \(\Psi(4660)\).

We can also use eq. \ref{eq:decaywidthboth} with \(A\) and \(\tilde A\) as determined from the pair \(\Psi(4360)/\Psi(4660)\) to calculate the width of the other states on the trajectory. However, given the ambiguities in defining the tension and in the question of whether or not to include of the phase space factor for higher states,\footnote{As was explained in \cite{Sonnenschein:2017ylo}, the factor \(\Phi\) is added by hand only in those cases where there is a single allowed channel with limited phase space. For higher excited states the growth of the total width should be simply linear in the string length.} we write only estimates. Unlike in \cite{Sonnenschein:2016ibx} we now use the two states rather than only \(Y(4360)\) as input. If we measure the slope between \(\Psi(4360)\) and \(\Psi(4660)\) we get \(\alp_n = 0.60 \GEVm\), exactly as for the charmonium trajectories.

The next state \(1^{--}\) should be just below 4900 MeV and decay predominantly to \(\Lambda_c\overline\Lambda_c\). We estimate the ratio \(\Gamma_{\text{tear}}/\Gamma_{\text{annihilation}}\) for it to be roughly between 3 and 6.

The \(\Psi(4360)\) and \(Y(4660)\) could also have a scalar state below them on their respective orbital HMRTs, with \(J^{PC}=0^{++}\). These states are calculated to be at 4170 and 4460 MeV. There is a candidate for the latter state, which is \(\chi_{c0}(4500)\) (also known as \(X(4500)\)). This state was seen in the \(J/\Psi \phi\) channel, and it is an exotic candidate of mass 4506\(^{+16}_{-19}\) MeV and width \(92\pm29\) MeV. This matches with the prediction in table \ref{tab:pred_y_c_j}.
%%%%%%%%%%%%%%%%%%%%%%%%%%%%%%%%%%%%%%%%%%%%%

\begin{table}[h!] \centering
	\begin{tabular}{|c|c|c|} \hline
\(n\)	&	 Mass 	&	 Width \\ \hline\hline
``-1'' & 4070 & 160--200\\ \hline\hline
0 & \(\bf{4368\pm13}\) & \(\bf{96\pm7}\)  \\ \hline
1 & \(\bf{4633\pm6}\)	&	\(\bf{64\pm9}\)	\\ \hline\hline
2 &	4870 	&	 80--210	\\ \hline
3	&	5100	&	 100--220	\\ \hline
4	&	5300 	&	 120--240	\\ \hline
\end{tabular}
\caption{\label{tab:pred_y_c_n} States on the radial trajectory of the \(\Psi(4360)\)/\(\Psi(4660)\). The higher excited states are expected to decay into \(\Lambda_c\overline{\Lambda}_c\). Another possible lower state, which is expected to be wider than \(\Psi(4360)\), is also included.}
\end{table}

\begin{table}[h!] \centering
\begin{tabular}{|c|c|c|} \hline
\(J^{PC}\) &	  Mass 	&	 Width \\ \hline\hline
\(0^{++}\) & 4170	& 160--200 \\ \hline\hline
\(1^{--}\) & \(\bf{4368\pm13}\) & \(\bf{96\pm7}\) 	\\ \hline\hline
\(2^{++}\) & 4550 &	50--80 	\\ \hline
\(3^{--}\) & 4720	& 70--170	 	\\ \hline
\(4^{++}\) & 4880 	&	80--210 	\\ \hline
\end{tabular} \qquad\qquad\qquad
\begin{tabular}{|c|c|c|} 
\hline
\(J^{PC}\) &	  Mass 	&	 Width \\ \hline\hline
\(0^{++}\) & 4460	&  70--100 \\ \hline\hline
\(1^{--}\) & \(\bf{4633\pm6}\)	&	\(\bf{64\pm9}\) 	\\ \hline\hline
\(2^{++}\) & 4800 	&	90--210 	\\ \hline
\(3^{--}\) & 4960	&	100--230 	\\ \hline
\(4^{++}\) & 5110	&	 120--240	\\ \hline
\end{tabular}
\caption{\label{tab:pred_y_c_j} Orbital trajectories of the \(\Psi(4360)\) and \(\Psi(4660)\). Here we use the slope of \(\alp_J=0.88\) \GEV\, to predict orbitally excited partners of the two states. If the states are tetraquarks there may be a lower state with \(J^{PC}=0^{++}\) as the ground state (which would not be the case for an ordinary \(c\bar c\)).}
\end{table}

\section{Analyzing the \texorpdfstring{$X(6900)$}{X(6900)} and \texorpdfstring{$X(7200)$}{X(7200)} exotic hadron states} \label{sec:cccc}

We use the \(\Psi(4360)\)/\(Y(4630)\)  system just described as our guiding line for the analysis of the newly discovered \(X(6900)\) and the unconfirmed state near 7.2 GeV, which we call \(X(7200)\). We also address the wider state near 6490 MeV.

We compute the states' masses based on an assignment to a modified Regge trajectory, and evaluate their widths within the HISH model.

\subsection{The structure and masses} \label{sec:ccccmasses}
We explore the system assuming it is a heavier analogue of the \(\Psi(4360)\)/\(Y(4630)\) system described in the previous section. The basic assumption is that it is a fully charmed \BV tetraquark, with the quark content \(cc\bar c\bar c\), which is built as a string connecting a diquark to an anti-diquark. The tetraquark made up of four heavy quarks has been an object of theoretical study for some decades, including models of diquark-anti-diquark bound states \cite{Karliner:2016zzc,Chao:1980dv,Lloyd:2003yc,Chen:2016jxd,Wu:2016vtq,Debastiani:2017msn,Wang:2017jtz,Liu:2019zuc,Chen:2020xwe,Giron:2020wpx}.
In the holographic realization this configuration involves a BV and an anti-BV.

In the HISH model the states are described by a string with massive particles on its endpoints, connecting a diquark of mass \(m_{cc}\approx 2m_c\), to an anti-diquark of the same mass.
%The state \(Y(4630)\) has a mass $M_Y = 2 M_{\Lambda_c} + \Delta M_c$, where $M_{\Lambda_c}$ is the mass of the lightest charmed baryon ${\Lambda_c}$, and  $\Delta M_c \approx 60$ MeV. Thus, the mass of the \(Y(4630)\) is very close to twice the mass of ${\Lambda_c}$.
% 
%In an analogous manner we conjecture that there should be an exotic tetraquark, with the structure depicted in figure 2 that has a mass in the vicinity of twice the mass of $\Xi_{cc}$, namely $M = 7242+ \Delta M_{cc} \MEV$.  At this point the question whether based  on the experimental value of $\Delta M_{c}$ we can estimate the value of $\Delta M_{cc}$.

%The next question is whether the mass difference $M_{Y'_{cc}}- M_{Y_{cc}}\simeq 7.2-6.9 Gev$ fits the mass different between adjecent states on the HMRT. Differently stating if we assume the ground state at $M_{Y_{cc}}= 6.9 Gev$ whether the first excited state is around $7.2 Gev$. To compute the theoretical HISH mass difference we use (\ref{eq:massGenJ}). Note that the intercept drops out of this computation. On the other hand it does depend on the slope $\alpha'$. For the radial excitation  the slope for the charmonia was found to be $\alp\!_n = 0.59$ and for the bottomonia $\alp\!_n = 0.46$.
%For these two values we get that the difference between the two states 
%$1.16>\Delta n >0.9$ which is in accordance with a picture of a ground state and a first excited state.

The experimental data are taken from \cite{Aaij:2020fnh}. There, only the mass of \(X(6900)\) is fully specified and measured. However, we can read from the fit done in their figure 7 where the other peaks are. There is a lower peak, wider than \(X(6900)\), centered at around 6490 MeV which is dubbed a threshold enhancement in \cite{Aaij:2020fnh}, rather than a resonance. The higher peak, the \(X(7200)\) is a Breit-Wigner resonance located around 7240 MeV. This is very near the threshold for baryon-antibaryon decay which is located at
\be 2M_{\Xi_{cc}} = 7242.4 \pm 1.4 \MEV \ee
The four peaks in figure 7 of \cite{Aaij:2020fnh} are measured by us to be centered at 6260, 6490, 6900, and 7240 MeV. See figure \ref{fig:peaks}.

%%%%%%%%%%%%%%%%%%%%%%%%%%%%%%%%
 
\subsubsection{The HISH modified Regge trajectories}
The tetraquarks that are built from a BV and and anti-BV all share two characteristics. The first is that, if they are heavy enough, their natural decay mode is to a baryon-antibaryon pair. The second property is that excited states, whether with higher angular momentum or with radial excitation number, should reside on HMRTs. These are based on the spectrum of the string connecting the BV and the anti-BV.

In \cite{Sonnenschein:2016ibx} we determined the trajectories associated with tetraquarks made up of diquarks that contain one $b$, $c$, or $s$ quark and another light quark, e.g. \(bq\bar b \bar q\) or \(cq\bar c\bar q\), where \(q\) is \(u\) or \(d\). The results for the charmonium-like \BV are given above in table 2.   In a similar manner we conjecture the trajectories on which the fully charm \(cc\bar c\bar c\) tetraquark resides.

The parameters which determine the HMRT of a given hadron are the endpoint masses, the string tension (equivalently the Regge slope), and the intercept.

For the \(cc\bar c \bar c\) state, we take a mass of \(m_{cc} = 2m_c\) for both the diquark and anti-diquark at the endpoint. The mass of the charm quark as a string endpoint was measured from the spectrum of HMRTs of charmed mesons \cite{Sonnenschein:2016ibx,Sonnenschein:2018fph}, and is \(m_c = 1490\) MeV.

The value of the Regge slope is also known, but we have found that while it is a universal parameter for light hadrons, it starts to depend on the mass when looking at \(c\bar c\) and heavier states. Since the \(X(6900)\) is in an intermediate range between \(c \bar c\) and \(b \bar b\) we could expect the relevant slope to be somewhere between those measured for the \(c\bar c\) and \(b\bar b\) trajectories.

We also observe a difference between the slopes of orbital and radial trajectories, which we denote by \(\alp_J\) and \(\alp_n\) respectively. The results of \cite{Sonnenschein:2018fph} are that for light mesons the slopes are
\be \text{light:} \qquad \alp_J = 0.88 \GEVm \qquad \alp_n = 0.80 \GEVm \ee
This includes the trajectories of all mesons containing only \(u\), \(d\) and \(s\) quarks, and also heavy-light mesons, i.e. those with a single \(c\) or \(b\). For the charmonium the orbital slope is the same, but the radial one is lower,
\be c\bar c: \qquad \alp_J = 0.88 \GEVm \qquad \alp_n = 0.60 \GEVm \ee
while for the bottomonium trajectories both are lower
\be b\bar b \qquad \alp_J = 0.55 \GEVm \qquad \alp_n = 0.42 \GEVm \ee
We also have a single measurement of the radial slope coming from a pair of \(B_c\) mesons, which gives \(\alp_n = 0.56 \GEVm\), a value between the charmonium and bottomonium slopes. We assume that the slope of the trajectory of the \BV state which is composed of \(cc\) and \(\bar c\bar c\) diquarks with mass \(2m_c\) will be in the same range, which is assuming that there is a smooth monotonous function of the mass that interpolates the values at \(m_c\) and \(m_b\).

To summarize, we assume the \(X(6900)\) belongs on a HMRT belonging to a string with masses \(m_{cc} \approx 2m_c\) on both of its endpoints, and an effective slope between the charmonium and bottomonium values. Then we calculate the intercept for each slope, and then the higher states on the trajectory. We can also go backwards on a trajectory and find where a lower, unexcited state would be. We can go at most two steps backward before we reach masses lower than \(4m_c\). For lower values of the slope the \(n=-2\) state is not present.

We predict the first few states on the trajectory of the \(X(6900)\) in table \ref{tab:pred_Ycc_n}. The mass of \(X(6900)\) is taken to be 6895 MeV for the purpose of the calculations, being a simple average over the two measurements. Experimental uncertainties should be added to the estimates in the table.

One result that is not dependent on the slope in this range is that the \(n=1\) state around 7200 MeV is generally expected to be below or very close to the threshold of two doubly charmed baryons. This is in contrast to the \(Y(4630)\), which is 60 MeV above threshold.

\begin{figure}[ht!] 
\centering
\includegraphics[width=0.66\textwidth]{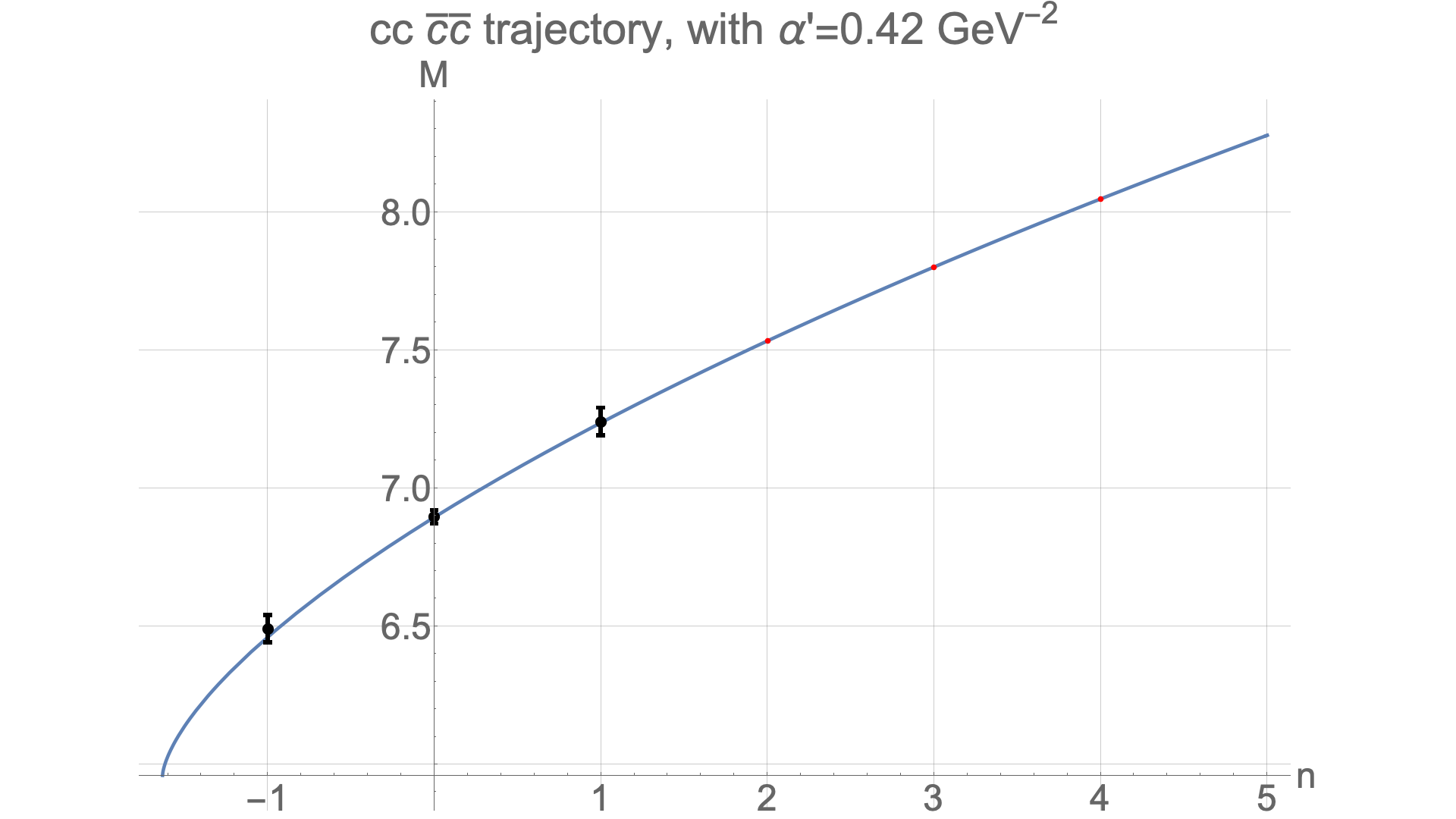}
\caption{\label{fig:6900spectrum} Radial HMRT of the \(X(6900)\) with \(\alp = \alp_{b\bar b} = 0.42\) \GEV. The \(X(6900)\) is placed at \(n=0\) although it might not be the ground state. The resonances at 6490 MeV at 7240 MeV, placed here at \(n=-1\) and \(1\) respectively, both fit the trajectory well. The next few states are at 7530, 7800, and 8050 MeV.}
\end{figure}

In fact, there appears to be good agreement between the Regge trajectory with the \(b\bar b\) slope, \(\alp_n = 0.42 \GEVm\), and the three states seen by LHCb. If we take as input the mass of \(X(6900)\) and that value of \(\alp_n\), then going backward on its trajectory we should find a state with mass 6470 MeV, and a higher state near 7240 MeV, which nearly exactly matches with \cite{Aaij:2020fnh}. See figure \ref{fig:6900spectrum}. On this trajectory the next state, which should decay predominantly to \(\Xi_{cc}\bar \Xi_{cc}\), would be near 7500 MeV.

\begin{table} \centering
\begin{tabular}{|c|c|} \hline
$n$ & Mass \\ \hline
``-2'' & 6000--6220 \\

``-1'' & 6460--6610 \\

0 & 6895 \\

1 & 7140--7230 \\

2 & 7350--7530 \\

3 & 7560--7800 \\
\hline
\end{tabular} \qquad\qquad\qquad
\begin{tabular}{|c|c|} \hline
$J_{orb}$ & Mass \\ \hline
``-2'' & 6110--6480 \\

``-1'' & 6570--6700 \\

0 & 6895 \\

1 & 7060--7160 \\

2 & 7220--7390 \\

3 & 7360--7610 \\
\hline
\end{tabular}
\caption{\label{tab:pred_Ycc_n} Trajectories of the \(X(6900)\), radial and orbital. We take values between 0.42 and 0.60 \GEV\,for the radial trajectory slope \(\alp_n\), while \(\alp_J\) goes between 0.55 and 0.88 \GEV. For lower slopes the mass differences between successive states are higher and the intercept smaller (in absolute value). The intercept is obtained to be between -2.3 and -1.6 for the radial trajectory and between -3.4 and -2.1 for the orbital one, depending on the slope.}
\end{table}

\subsection{Decays of the \texorpdfstring{$X(6900)$}{X(6900)} and \texorpdfstring{$X(7200)$}{X(7200)} states } \label{sec:ccccdecay}
As explained in section \ref{sec:decays} there are different mechanisms of decay for states below or above the baryon-antibaryon threshold, as we have argued is the case for the pair $\Psi(4360)$ and $Y(4630)$. Here we examine the states $X(6900)$ and $X(7200)$, in direct analogy to the discussion in section \ref{sec:cucu}.

The states below the baryon-antibaryon threshold are expected to decay in a process that involves annihilation of the BV and anti-BV, a process which is suppressed exponentially in the string length. Above threshold, they should decay via tearing of the string, which is proportional to the length. States above but close to the threshold are expected to have both modes.

\subsubsection{The decay of the \texorpdfstring{$X(6900)$}{X(6900)}}
The mechanism of the decay of the $X(6900)$ was explained in \ref{sec:31}. The probability of the decay process is given in eq. \ref{eq:decaywidthannihilation}.
To avoid a complicated calculation of the pair of BV anti-BV annihilation we will make the plausible assumption  that it is the same for the $\Psi(4360)$ system and the $X(6900)$ and hence the ratio of the widths of the latter to the former is
\be
\frac{\Gamma_{X(6900)}}{\Gamma_{\Psi(4360)}} \simeq \frac{\exp\left(-\frac{TL^2}{2}|_{X(6900)}\right)}{\exp\left(-\frac{TL^2}{2}|_{\Psi(4360)}\right)}\approx \frac{\exp\left(-\frac{4(M_{X(6900)}-4m_c)^2}{9 T}\right)}{\exp\left(-\frac{4(M_{\Psi(4360)}-2(m_c+m_d))^2}{9 T}\right)}  
\ee
Since it is a ratio of two exponential terms it is quite sensitive to uncertainties of the parameters and we cannot determine an accurate prediction of the width. In addition, there is no guarantee that the states share the same string tension, so we must account for that as well. By checking some plausible values of the string tension and the diquark masses, we see that the ratio is always larger than \(\approx 1.9\). Since the width of \(\Psi(4360)\) is \(96\pm7\) MeV, the model suggests that
\be \Gamma[X(6900)] \gtrsim 180\,\MEV \ee
Therefore it is more compatible with the larger width of \(168\pm33\pm69\) MeV for \(X(6900)\).

We can also use the formula to predict the ratio of widths between \(X(6900)\) and the lower state at 6490 MeV, assuming that it shares the same \BV structure, with the result that is should be wider than \(X(6900)\) by a factor of 1.4--2.0. This qualitatively matches the data, which shows that \(X(6900)\) is the narrower state. Based on the lower bound above, the 6490 MeV state should have a width of at least 250 MeV.

Note that the state at 6900 MeV is heavy enough to also decay to \(J/\Psi D \bar D\), which would be interesting to see in experiment.\footnote{We thank Ivan Polyakov for pointing this out to us.} This can also lower the prediction for the width of the lower state depending on the branching ratio.

\subsubsection {The decay of the \texorpdfstring{$X(7200)$}{X(7200)}}
It is not clear whether the \(X(7200)\) is heavy enough for the baryon-antibaryon decay to \(\Xi_{cc}\bar \Xi_{cc}\) to occur. The data from LHCb suggests that it is very near to the threshold, and so do our calculations of its mass as the first radially excited partner of the \(X(6900)\).

If it is above threshold the natural decay mechanism, as was shown in figure \ref{fig:decaysf}, is of breaking the string and creating a quark and antiquark. The decay products are then a baryon-antibaryon pair. 

We can estimate the decay width of the \(X(7200)\) in two possible models. The first is assuming that it is above threshold and decays only to the baryon-antibaryon, and assuming that likewise \(Y(4630)\) decays only to \(\Lambda_c\overline\Lambda_c\). Then we can estimate that the ratio of the two states' widths is approximately given by the ratio of their lengths, adjusted for phase space corrections:
\be
\frac{\Gamma_{X(7200)}}{\Gamma_{Y(4630)}}\simeq \frac{(TL)[X(7200)]}{(TL)[Y(4630)]}\times  \frac{\Phi[X(7200)\to \Xi_{cc}\bar\Xi_{cc}]}{\Phi[Y(4630)\to \Lambda_c\bar \Lambda_c]} 
\ee
This can be written solely as a function of the mass of the \(X(7200)\), or in terms of \(\Delta M = M_{X(7200)}-2M_{\Xi_{cc}}\). It is given by
%\be \frac{\Gamma_{X(7200)}}{\Gamma_{Y(4630)}} \approx \frac{2M_{\Xi_{cc}}-4m_c + \Delta M}{M_{Y(4630)}-2m_c} \frac{\sqrt{1-\frac{4M_{\Xi_{cc}}^2}{(2M_{\Xi_{cc}}+\Delta M)^2}}}{\sqrt{1-\frac{4M_{\Lambda_c}^2}{M_{Y(4630)}^2}}} \ee
\be \frac{\Gamma_{X(7200)}}{\Gamma_{Y(4630)}} \approx 4.78 \sqrt x(1+2.45 x) \label{eq:Gamma7200} \ee
where \(x = \frac{\Delta M}{M_{\Xi_{cc}}}\). We can use this simple formula to estimate the decay width. For instance, if the state is just above threshold, with \(\Delta M = 10 \MEV\), then $\Gamma_{X(7200)} \approx 23 \MEV$. For a larger mass difference $\Delta M= 50 \MEV$ we get that $\Gamma_{X(7200)} \approx 53 \MEV$, and so on. Given the large error in the measured width of \(Y(4630)\) from \(\Lambda_c\bar\Lambda_c\) (eq. 4.1), then these values will also have an error of about 40--50\% in addition to uncertainties from the model.

The second model is more in line with what was presented in section \ref{sec:cucu}, and includes the decays of \(X(7200)\) by BV-anti-BV annihilation, resulting in a pair of charmonia. In the same way we calculated the width of \(\Psi(4660)\) based on the width of the below threshold \(\Psi(4360)\), we can use the \(X(6900)\) to give an estimate of \(\Gamma[X(7200]\) when \(X(7200)\) is below threshold. Then its width should be narrower than that of \(X(6900)\) as given by the ratio of \(\exp(-TL^2/2)\), as in the last section. The result is that the width of the \(X(7200)\) should be 45--70\% of the width of the \(X(6900)\), depending on the state's exact mass and the value used for the tension.

Then there is the possibility that the state is above threshold and both decay channels are present, as we argued was the case for the \(Y(4630)/\Psi(4660)\). In that case eq. \ref{eq:Gamma7200} becomes the ratio of the partial decays widths \(\Gamma_{\text{tear}}[X(7200)]/\Gamma_{\text{tear}}[\Psi(4660)]\).

Then the decay width is estimated as
\be \Gamma[X(7200)] \approx (0.45\text{--} 0.70)\times \Gamma[(X(6900)] + 4.78 \sqrt{x} \times \Gamma_{\text{tear}}[\Psi(4660)] \ee
where \(\Gamma_{\text{tear}}[\Psi(4660)]\) was about 20--30 MeV. The second term is less than 10 MeV as long as \(\Delta M \lesssim 40\) MeV, so the larger part will come from the first term, which is in the range 40--140 MeV depending on the measurement used for \(\Gamma[X(6900)]\).

%%%%%%%%%%%%%%%%%%%%%%%%%%%%%%%%%%%%%
\section{Predictions about novel \texorpdfstring{\BV}{V-baryonium} tetraquarks} \label{sec:novelbv}
%%%%%%%%%%%%%%%%%%%%%%%%
Using the same logic, as we have used above it is natural to predict other \BV states. For example an exotic tetraquark built of diquark of $c$ and $s$ quarks with mass
$M_{Y_{cs}} =2 M_{\Xi_c}+ \Delta M_{cs} = 4.936\text{ GeV }+ \Delta M_{cs}$, where  $\Delta M_{cs}$ is the difference between the actual mass of the hadron and the threshold mass which is sum of mass of the baryon and anti-baryon. For the case $Y(4630)$, $\Delta M\approx 60$ MeV.

The  states conjectured to have a structure of a tetraquark -  $Y(4630)$, $X(6900)$ and related states discussed above, and $Y_{cs}$ just discussed are symmetric in their structure in the sense that for the quarks $q_1$, $q_2$ attached to the baryonic vertex, on the other side there are $\bar q_1$ and $\bar q_2$ which are attached to the anti-BV.

However, as was mentioned already in \cite{Sonnenschein:2016ibx}, there is no reason not to expect also asymmetric tetraquarks, namely where on the anti-BV there will be $\bar q_3$ which is different than $\bar q_1$ and $\bar q_2$ and similarly also for $\bar q_4$. For exotics that include both $c$ and $\bar c$ asymmetric tetraquarks can have the following structures
$(c, c,\bar c, \bar s)$,  $(c, c,\bar c, \bar q)$,  $(c, s,\bar c, \bar q)$, where \(q=u,d\).
If these exotic hadrons admit the structure described above their masses should be somewhat larger than

\[ M_{Y_{c,c,\bar c,\bar s}}\gtrsim M_{\Xi_{cc}}+ M_{\Xi_c}=6089 \text{ MeV}\]

\[ M_{Y_{c,c,\bar c,\bar q}}\gtrsim M_{\Xi_{cc}}+ M_{\Lambda_c}=5897 \text{ MeV}\]

\[ M_{Y_{c,s,\bar c,\bar q}}\gtrsim M_{\Xi_{c}}+ M_{\Lambda_c}=4754 \text{ MeV}\]

%%%%%%%%%%%%%%%%%%%%%%%%%%%
In \cite{Sonnenschein:2016ibx} we have presented a detailed analysis of the analogs of the $Y(4630)$ for the exotics build from strange and bottom quarks, namely tetraquarks where there is one bottom/strange quark attached to the BV and a single  anti-quark with the same flavor to the anti- BV. In analogy to the passage from a charmonium-like to  the fully charm \BV there may be exotics related to such transformations also for the bottom, strange, and possibly for light-quarks of the form $Y_q\rightarrow Y_{qq}$ where $q$ stands for $b$, $s$, $u/d$.

In table \ref{tab:others} we list the threshold masses for all possible  symmetric \BV. In the first two lines, since the corresponding baryons have not be found yet, we estimated their masses instead. In all other lines we use the PDG values of the corresponding baryons.
% This generalization can be for exotics which are symmetric under the interchange between the BV and anti-BV and hence are charge-less.
 
% In the bottom sector it includes the following contents and threshold masses: 
%$$(b, b , \bar b,\bar b)\ ?  \qquad (b, c , \bar b,\bar c)\ ?  \qquad
%(b, s , \bar b,\bar s)\  11594\qquad
%(b, u/d , \bar b,\bar u/d)\ 11240,$$ 
% There could be also asymmetric exotics in that sector with diquark of the same flavor  
%$$(b, b , \bar c,\bar c)\ ? \qquad (b, b , \bar s,\bar s)\ ?  \qquad
%(b, b , \bar u/d,\bar u/d)\  ? ,$$ 
%or of different flavor with (b,b) on the BV side 
%$$(b, b , \bar c,\bar s)\qquad (b, b , \bar c,\bar u/d) \qquad
%(b, b , \bar s,\bar u/d),$$
%or with (b,c) on that side
%$$(b, c , \bar b,\bar s)\qquad (b, c , \bar b,\bar u/d) \qquad
%(b, c , \bar c,\bar s)\qquad
%(b, c , \bar c,\bar u/d)\qquad
%(b, c , \bar s,\bar u/d),$$
%or (b,s) on that side
%$$(b, s , \bar b,\bar u/d)\ 11417\qquad (b, s , \bar c,\bar s)\ 8265 \qquad
%(b, s , \bar c,\bar u/d)\ 7983 \qquad
%(b, s , \bar s,\bar u/d)\ 6913.$$
%
%Possibly there could be such resonances also in the strange sector.
%The symmetric   $s$ sector  includes 
%$$(s, s , \bar s,\bar s)\ 2628 \qquad (s, u/d , \bar s,\bar {u/d}) \ 2232 $$ and asymmetric strange ones
%$$(s, s , \bar s,\bar u/d)\ 2430 \qquad (s, s , \bar {u/d},\bar {u/d}) \ 2252 \qquad
%(s, u/d , \bar u/d,\bar u/d)\ 2054,$$
%%%%%%%%%%%%%%%%%%%%%%%%%%%%%%%%%%%%%%%%%

\begin{table} \centering
\begin{tabular}{|c|cc|c|} \hline
Quark & \multicolumn{2}{|c|}{Decay products} & Threshold \\
content &  \multicolumn{2}{|c|}{(Above threshold)} & [MeV] \\ \hline\hline

$b b \bar b \bar b$  & [$\Xi_{bb}$] & [$\bar\Xi_{bb}$] & $\sim$ 20600 \\ \hline
$b c \bar b \bar c$  & [$\Xi_{bc}$] & [$\bar\Xi_{bc}$] &$\sim$ 14200 \\ \hline
$b s \bar b \bar s$  &$\Xi_b$ & $\bar\xi_b $ & 11594 \\ \hline
$b q \bar b \bar q$  &$\Lambda_b$ & $\bar\Lambda_b$ & 11240 \\ \hline
$c c \bar c \bar c$ &$\Xi_{cc}$ & $\Xi_{cc}$ & 7240 \\ \hline
$c s \bar c \bar s$ &$\Xi_c$ & $\bar\Xi_c$ & 4936 \\ \hline
$c q \bar c \bar q$ &$\Lambda_c$ & $\Lambda_c$ & 4572 \\ \hline

$s s \bar s \bar s$ & $\Xi$ & $\bar\Xi$ & 2628 \\ \hline
$s q \bar s \bar q$ &$\Lambda$ & $\bar\Lambda$ & 2232 \\ \hline
$q q \bar q \bar q$  &$ N$ & $\bar N$ & 1876 \\ \hline

\end{tabular} \caption{\label{tab:others} Potential symmetric (flavorless) \BV states and the thresholds for their baryon-antibaryon decays. \(q\) stands for $u$ or $d$. The doubly bottom \(\Xi_{bb}\) and bottom-charm \(\Xi_{bc}\) baryons are currently still unobserved.}
\end{table}

%\begin{table} \centering
%\begin{tabular}{|c|c|c|c|cc|c|} \hline
%Quark & \(B\) & \(C\) & \(S\) & \multicolumn{2}{|c|}{Decay products} & Threshold \\
%content & & & & \multicolumn{2}{|c|}{(Above threshold)} & [MeV] \\ \hline\hline
%
%$b b \bar b \bar b$ & 0 & 0 & 0  & [$\Xi_{bb}$] & [$\bar\Xi_{bb}$] & $\sim$ 20600 \\ \hline
%$b c \bar b \bar c$ & 0 & 0 & 0  & [$\Xi_{bc}$] & [$\bar\Xi_{bc}$] &$\sim$ 14200 \\ \hline
%$b s \bar b \bar s$ & 0 & 0 & 0  &$\Xi_b$ & $\bar\xi_b $ & 11594 \\ \hline
%$b q \bar b \bar q$ & 0 & 0 & 0  &$\Lambda_b$ & $\bar\Lambda_b$ & 11240 \\ \hline
%$c c \bar c \bar c$ & 0 & 0 & 0  &$\Xi_{cc}$ & $\Xi_{cc}$ & 7240 \\ \hline
%$c s \bar c \bar s$ & 0 & 0 & 0  &$\Xi_c$ & $\bar\Xi_c$ & 4936 \\ \hline
%$c q \bar c \bar q$ & 0 & 0 & 0  &$\Lambda_c$ & $\Lambda_c$ & 4572 \\ \hline
%
%$s s \bar s \bar s$ & 0 & 0 & 0  & $\Xi$ & $\bar\Xi$ & 2628 \\ \hline
%$s q \bar s \bar q$ & 0 & 0 & 0  &$\Lambda$ & $\bar\Lambda$ & 2232 \\ \hline
%$q q \bar q \bar q$ & 0 & 0 & 0  &$ N$ & $\bar N$ & 1876 \\ \hline
%
%
%\end{tabular} \caption{Potential  symmetric \BV states and the thresholds for their baryon-antibaryon decays. \(B\), \(C\), and \(S\) denote flavor (bottomness, charm, and strangeness respectively).q stands for $u$ or $d$. The doubly bottom \(\Xi_{bb}\) and bottom-charm \(\Xi_{bc}\) baryons are currently still unobserved.}
%\end{table}

It is trivial to determine the threshold masses for all of them. For a \BV with a content of $(q_1,q_2,\bar q_3,\bar q_4)$ the threshold mass is the sum of the masses of the lightest baryon $(q_1,q_2,q)$ and the anti-baryon $(\bar q_1,\bar q_2,\bar q)$ where $q$ is $u$ or $d$.

$$ M_{Y_{(q_1,q_2,\bar q_3,\bar q_4)}}\gtrsim M_{B_{(q_1,q_2,q)}}+ M_{\bar B_{(\bar q_3,\bar q_4,\bar q)}}$$

In all the decays described in the table we have assumed that the pair created by the breaking of the string is a pair of light quarks. In general, if the mass of the tetraquark is high enough heavier pair can also be created. In particular a pair of $(s,\bar s)$. In \cite{Sonnenschein:2017ylo} the ratio of the widths of decays involving a creation of $(s,\bar s)$ over that of  $(u/d,\bar u/\bar d)$ was found out to be $\sim 0.3$. For example the excited state with $n=4$ in table \ref{tab:pred_y_c_n} can decay not only to $\Lambda_c, \bar \Lambda_c$ but also to $\Omega_c,\bar\Omega_c$. 

The search for the \BV states conjectured in this section  should be based on identifying states that decay to a baryon and an anti-baryon. If fact we would like to argue that this search may also lead to finding of baryons that have not been detected yet \cite{Karliner:2019vhw,Karliner:2019hbm} like those of $ (bcs), (bcq), (bbs), (bbq)$. The fully heavy baryons as $ (ccc), (bcc), (bbb)$ are unlikely to be found from decays of \BV, since those would require pair creation of heavy quarks. 
 
\section{Summary} \label{sec:summary}
In this note we studied the system of tetraquark candidates around  $X(6900)$ in analogy to the system that includes the $\Psi(4360)$ and $Y(4630)$/\(\Psi(4660)\). The latter was conjectured in \cite{Sonnenschein:2016ibx} to be a system of \BV tetraquarks, which we argue is the case also for the new states.

The smoking gun for this conjecture would be to observe higher excited states predominantly decay into a baryon and anti-baryon.  In this note, complementing the discussion of \cite{Sonnenschein:2016ibx}, we account also for a decay mechanism involving annihilation of the BV and anti-BV in the tetraquark, for the ground and other lower states that are too light to decay to the baryon-anti-baryon pair, or have limited phase space to do so.

Another signature of this conjecture is the HISH modified Regge trajectories associated with each \BV state. Using the HMRTs we can predict the masses of excited states, both orbital and radial excitations. The excited \BV tetraquarks would predominantly decay to a baryon-antibaryon pair, and we would like to encourage experiments looking at these channels.

The proposed upgrade of BEPCII, which will allow the BESIII experiment it to access energies up to 4.9 GeV is a perfect opportunity to revisit the \(Y(4630)/\Psi(4660)\) state in the \(\Lambda_c\bar\Lambda_c\) channel \cite{Ablikim:2019hff}. Initial measurements near the threshold reveal some discrepancies with Belle data \cite{Ablikim:2017lct}. Furthermore, we predict an excited \(J^{PC} = 1^{--}\) state at around 4870 MeV, which should decay predominantly to \(\Lambda_c\bar\Lambda_c\), which is still in that range. The threshold of \(\Lambda_c\Sigma_c\) at 4.74 GeV will also be accessible, so one could potentially discover also a tetraquark state that decays to those baryons.

The \(X(6900)\) opens a new range for future experiments \cite{Wang:2020gmd,Wang:2020wrp}. The unconfirmed second resonance \(X(7200)\) is at the right mass to be a radial excitation of \(X(6900)\). The lower, wider resonance around 6490 MeV also fits the same HMRT. The \(X(7200)\) is interesting being very near the threshold for decay to \(\Xi_{cc}\) and \(\overline{\Xi}_{cc}\). A confirmation and accurate measurement of its properties will be important in understanding the system. The next excitation, which would decay predominantly to the doubly charmed baryon pair, is predicted at around 7500 MeV. It should not be much wider than \(X(7200)\).

There are also many open questions regarding tetraquarks and the stringy model of the \BV. Let us list some here.
\begin{itemize}
\item
A natural question about the HISH construction of exotics, is what about exotic hadrons other than tetraquarks, such pentaquarks, hexaquarks, and so on. In \cite{Sonnenschein:2016ibx} we schematically described other stringy constructions of exotic hadrons. The quest for more complicated exotics requires further investigation. For example one can construct an exotic hexaquark using three BV with three diquarks, all connected through an anti-BV such that in total we get an object of baryon number 2 and with six quark endpoints. This object may related to the exotic hadron discussed in \cite{Azizi:2019xla,Farrar:2020zeo}.
\item
Whereas quarks and strings (flux tubes) are well understood in QCD the BV is much less familiar. In holography it is described as a $D_0$ brane or a fractional $D_0$ brane, namely a $D_p$ brane wrapping a $p$-cycle. It carries the baryon number charge. Other properties like its mass, the structure of the diquark as a BV and two short strings, or the annihilation process of BV with anti-BV all deserve further study. In QCD it is also referred to as a string junction (see \cite{Rossi:2016szw} for a recent review), and is an important ingredient in some phenomenological models, e.g. in \cite{Karliner:2016zzc}.
\item
One of the major open questions regarding exotics regards the existence of tetraquarks made up of light quarks only. It is probable that exotic tetraquarks are simply more stable for hadrons that include heavy flavor quarks. In the string model, the decay mechanism of  breaking up of the string is not very sensitive to whether the endpoints are light or heavy. However the other decay channel via annihilation of the BV and anti-BV may have higher probability when there are light diquarks attached to the annihilating pair. By this argument the \BV made up solely of light quarks would be wide and hence difficult to detect. This explanation deserves further investigation.

\item Very recently, an open flavor tetraquark candidate has been observed \cite{Aaij:2020ypa,Aaij:2020hon,Karliner:2020vsi}. These are two states with spins 0 and 1, both with a mass of around 2.9 GeV, and with the quark content \(cs\bar u \bar d\). Since the HISH model describes hadrons of both light and heavy quarks, this tetraquark could also be described using the same stringy model used in this note. For this system the relevant baryon-antibaryon threshold is of \(\Xi_{c}^+ \bar p\) or \(\Xi_{c}^0\bar n\) at 3400 MeV.

\end{itemize}
\section*{Acknowledgments}
We would like to thank Marek Karliner for informing us about the new discovery and to Avner Soffer for useful discussions.   This work was supported in part by a center of excellence supported by the Israel Science Foundation (grant number 2289/18).  DW is supported by the Quantum Gravity Unit at the Okinawa Institute of Science and Technology Graduate University.

\bibliographystyle{JHEP20}
\bibliography{novelexotics}

\providecommand{\href}[2]{#2}\begingroup\raggedright\begin{thebibliography}{10}

\bibitem{Aaij:2020fnh}
{\bf LHCb} Collaboration, R.~Aaij et~al., {\it {Observation of structure in the
  $J/\Psi$-pair mass spectrum}},  \href{https://arxiv.org/abs/2006.16957}{{\tt
  arXiv:2006.16957}}.

\bibitem{Sonnenschein:2018fph}
J.~Sonnenschein and D.~Weissman, {\it {Excited mesons, baryons, glueballs and
  tetraquarks: Predictions of the Holography Inspired Stringy Hadron model}},
  {\em Eur. Phys. J. C} {\bf 79} (2019), no.~4 326,
  [\href{https://arxiv.org/abs/1812.01619}{{\tt arXiv:1812.01619}}].

\bibitem{Sonnenschein:2016pim}
J.~Sonnenschein, {\it {Holography Inspired Stringy Hadrons}},  {\em Prog. Part.
  Nucl. Phys.} {\bf 92} (2017) 1--49,
  [\href{https://arxiv.org/abs/1602.00704}{{\tt arXiv:1602.00704}}].

\bibitem{Liu:2019mxw}
Y.~Liu, M.~A. Nowak, and I.~Zahed, {\it {Heavy Holographic Exotics: Tetraquarks
  as Efimov States}},  {\em Phys. Rev. D} {\bf 100} (2019), no.~12 126023,
  [\href{https://arxiv.org/abs/1904.05189}{{\tt arXiv:1904.05189}}].

\bibitem{Liu:2019yye}
Y.~Liu, M.~A. Nowak, and I.~Zahed, {\it {Heavy tetraquark $QQ\bar{q}\bar{q}$ as
  a hadronic Efimov state}},  \href{https://arxiv.org/abs/1909.02497}{{\tt
  arXiv:1909.02497}}.

\bibitem{Witten:1998xy}
E.~Witten, {\it {Baryons and branes in anti-de Sitter space}},  {\em JHEP} {\bf
  9807} (1998) 006, [\href{https://arxiv.org/abs/hep-th/9805112}{{\tt
  hep-th/9805112}}].

\bibitem{Rossi:1977cy}
G.~C. Rossi and G.~Veneziano, {\it {A Possible Description of Baryon Dynamics
  in Dual and Gauge Theories}},  {\em Nucl. Phys.} {\bf B123} (1977) 507--545.

\bibitem{Montanet:1980te}
L.~Montanet, G.~C. Rossi, and G.~Veneziano, {\it {Baryonium Physics}},  {\em
  Phys. Rept.} {\bf 63} (1980) 149--222.

\bibitem{Rossi:2016szw}
G.~Rossi and G.~Veneziano, {\it {The string-junction picture of multiquark
  states: an update}},  {\em JHEP} {\bf 06} (2016) 041,
  [\href{https://arxiv.org/abs/1603.05830}{{\tt arXiv:1603.05830}}].

\bibitem{Sonnenschein:2016ibx}
J.~Sonnenschein and D.~Weissman, {\it {A tetraquark or not a tetraquark? A
  holography inspired stringy hadron (HISH) perspective}},  {\em Nucl. Phys. B}
  {\bf 920} (2017) 319--344, [\href{https://arxiv.org/abs/1606.02732}{{\tt
  arXiv:1606.02732}}].

\bibitem{Pakhlova:2008vn}
{\bf Belle} Collaboration, G.~Pakhlova et~al., {\it {Observation of a
  near-threshold enhancement in the e+e- ---> Lambda+(c) Lambda-(c) cross
  section using initial-state radiation}},  {\em Phys. Rev. Lett.} {\bf 101}
  (2008) 172001, [\href{https://arxiv.org/abs/0807.4458}{{\tt
  arXiv:0807.4458}}].

\bibitem{Sonnenschein:2014bia}
J.~Sonnenschein and D.~Weissman, {\it {A rotating string model versus baryon
  spectra}},  {\em JHEP} {\bf 02} (2015) 147,
  [\href{https://arxiv.org/abs/1408.0763}{{\tt arXiv:1408.0763}}].

\bibitem{Sonnenschein:2014jwa}
J.~Sonnenschein and D.~Weissman, {\it {Rotating strings confronting PDG
  mesons}},  {\em JHEP} {\bf 08} (2014) 013,
  [\href{https://arxiv.org/abs/1402.5603}{{\tt arXiv:1402.5603}}].

\bibitem{Sonnenschein:2018aqf}
J.~Sonnenschein and D.~Weissman, {\it {Quantizing the rotating string with
  massive endpoints}},  {\em JHEP} {\bf 06} (2018) 148,
  [\href{https://arxiv.org/abs/1801.00798}{{\tt arXiv:1801.00798}}].

\bibitem{Sonnenschein:2017ylo}
J.~Sonnenschein and D.~Weissman, {\it {The decay width of stringy hadrons}},
  {\em Nucl. Phys. B} {\bf 927} (2018) 368--454,
  [\href{https://arxiv.org/abs/1705.10329}{{\tt arXiv:1705.10329}}].

\bibitem{Aaij:2020ypa}
{\bf LHCb} Collaboration, R.~Aaij et~al., {\it {Amplitude analysis of the
  $B^+\to D^+D^-K^+$ decay}},  \href{https://arxiv.org/abs/2009.00026}{{\tt
  arXiv:2009.00026}}.

\bibitem{Aaij:2020hon}
{\bf LHCb} Collaboration, R.~Aaij et~al., {\it {A model-independent study of
  resonant structure in $B^+\to D^+D^-K^+$ decays}},
  \href{https://arxiv.org/abs/2009.00025}{{\tt arXiv:2009.00025}}.

\bibitem{Karliner:2020vsi}
M.~Karliner and J.~L. Rosner, {\it {First exotic hadron with open heavy flavor:
  $cs\bar u\bar d$ tetraquark}},  \href{https://arxiv.org/abs/2008.05993}{{\tt
  arXiv:2008.05993}}.

\bibitem{Ali:2017jda}
A.~Ali, J.~S. Lange, and S.~Stone, {\it {Exotics: Heavy Pentaquarks and
  Tetraquarks}},  {\em Prog. Part. Nucl. Phys.} {\bf 97} (2017) 123--198,
  [\href{https://arxiv.org/abs/1706.00610}{{\tt arXiv:1706.00610}}].

\bibitem{Yuan:2018inv}
C.-Z. Yuan, {\it {The XYZ states revisited}},  {\em Int. J. Mod. Phys. A} {\bf
  33} (2018), no.~21 1830018, [\href{https://arxiv.org/abs/1808.01570}{{\tt
  arXiv:1808.01570}}].

\bibitem{Brambilla:2019esw}
N.~Brambilla, S.~Eidelman, C.~Hanhart, A.~Nefediev, C.-P. Shen, C.~E. Thomas,
  A.~Vairo, and C.-Z. Yuan, {\it {The $XYZ$ states: experimental and
  theoretical status and perspectives}},
  \href{https://arxiv.org/abs/1907.07583}{{\tt arXiv:1907.07583}}.

\bibitem{Cotugno:2009ys}
G.~Cotugno, R.~Faccini, A.~D. Polosa, and C.~Sabelli, {\it {Charmed
  Baryonium}},  {\em Phys. Rev. Lett.} {\bf 104} (2010) 132005,
  [\href{https://arxiv.org/abs/0911.2178}{{\tt arXiv:0911.2178}}].

\bibitem{Guo:2010tk}
F.-K. Guo, J.~Haidenbauer, C.~Hanhart, and U.-G. Meissner, {\it {Reconciling
  the X(4630) with the Y(4660)}},  {\em Phys. Rev.} {\bf D82} (2010) 094008,
  [\href{https://arxiv.org/abs/1005.2055}{{\tt arXiv:1005.2055}}].

\bibitem{Brodsky:2014xia}
S.~J. Brodsky, D.~S. Hwang, and R.~F. Lebed, {\it {Dynamical Picture for the
  Formation and Decay of the Exotic XYZ Mesons}},  {\em Phys. Rev. Lett.} {\bf
  113} (2014), no.~11 112001, [\href{https://arxiv.org/abs/1406.7281}{{\tt
  arXiv:1406.7281}}].

\bibitem{Liu:2016sip}
X.~Liu, H.-W. Ke, X.~Liu, and X.-Q. Li, {\it {Exploring open-charm decay mode
  $\Lambda _c\bar{\Lambda }_c$ of charmonium-like state $Y(4630)$}},  {\em Eur.
  Phys. J.} {\bf C76} (2016), no.~10 549,
  [\href{https://arxiv.org/abs/1601.00762}{{\tt arXiv:1601.00762}}].

\bibitem{Guo:2016iej}
X.-D. Guo, D.-Y. Chen, H.-W. Ke, X.~Liu, and X.-Q. Li, {\it {Study on the rare
  decays of $Y(4630)$ induced by final state interactions}},  {\em Phys. Rev.}
  {\bf D93} (2016), no.~5 054009, [\href{https://arxiv.org/abs/1602.02222}{{\tt
  arXiv:1602.02222}}].

\bibitem{PDG:2020}
P.~Zyla et~al., {\it {Review of Particle Physics}},  {\em PTEP} {\bf 2020}
  (2020), no.~8 083C01.

\bibitem{PDG:Live}
``{2020 Review of Particle Physics, accessed via PDG Live website}.''
  \url{https://pdglive.lbl.gov}.
\newblock Accessed: 2020-07-27.

\bibitem{Ebert:2008kb}
D.~Ebert, R.~N. Faustov, and V.~O. Galkin, {\it {Excited heavy tetraquarks with
  hidden charm}},  {\em Eur. Phys. J.} {\bf C58} (2008) 399--405,
  [\href{https://arxiv.org/abs/0808.3912}{{\tt arXiv:0808.3912}}].

\bibitem{Ding:2007rg}
G.-J. Ding, J.-J. Zhu, and M.-L. Yan, {\it {Canonical Charmonium Interpretation
  for Y(4360) and Y(4660)}},  {\em Phys. Rev.} {\bf D77} (2008) 014033,
  [\href{https://arxiv.org/abs/0708.3712}{{\tt arXiv:0708.3712}}].

\bibitem{Wang:2016fhj}
Y.-Y. Wang, Q.-F. Lü, E.~Wang, and D.-M. li, {\it {Role of $Y(4630)$ in the
  $p\bar{p}\rightarrow\Lambda_c\bar{\Lambda}_c$ reaction near threshold}},
  {\em Phys. Rev.} {\bf D94} (2016) 014025,
  [\href{https://arxiv.org/abs/1604.01553}{{\tt arXiv:1604.01553}}].

\bibitem{Anwar:2018sol}
M.~N. Anwar, J.~Ferretti, and E.~Santopinto, {\it {Spectroscopy of the
  hidden-charm $[qc][\bar q \bar c]$ and $[sc][\bar s \bar c]$ tetraquarks in
  the relativized diquark model}},  {\em Phys. Rev. D} {\bf 98} (2018), no.~9
  094015, [\href{https://arxiv.org/abs/1805.06276}{{\tt arXiv:1805.06276}}].

\bibitem{Sundu:2018toi}
H.~Sundu, S.~Agaev, and K.~Azizi, {\it {Resonance $Y(4660)$ as a vector
  tetraquark and its strong decay channels}},  {\em Phys. Rev. D} {\bf 98}
  (2018), no.~5 054021, [\href{https://arxiv.org/abs/1805.04705}{{\tt
  arXiv:1805.04705}}].

\bibitem{Yan:2018gik}
X.~Yan, B.~Zhong, and R.~Zhu, {\it {Doubly charmed tetraquarks in a
  diquark--antidiquark model}},  {\em Int. J. Mod. Phys. A} {\bf 33} (2018),
  no.~16 1850096, [\href{https://arxiv.org/abs/1804.06761}{{\tt
  arXiv:1804.06761}}].

\bibitem{Wang:2018rfw}
Z.-G. Wang, {\it {Vector tetraquark state candidates: $Y(4260/4220)$,
  $Y(4360/4320)$, $Y(4390)$ and $Y(4660/4630)$}},  {\em Eur. Phys. J. C} {\bf
  78} (2018), no.~6 518, [\href{https://arxiv.org/abs/1803.05749}{{\tt
  arXiv:1803.05749}}].

\bibitem{Wang:2019iaa}
Z.-G. Wang, {\it {Strong decays of the $Y(4660)$ as a vector tetraquark state
  in solid quark-hadron duality}},  {\em Eur. Phys. J. C} {\bf 79} (2019),
  no.~3 184, [\href{https://arxiv.org/abs/1901.02177}{{\tt arXiv:1901.02177}}].

\bibitem{Cao:2019wwt}
Q.-F. Cao, H.-R. Qi, Y.-F. Wang, and H.-Q. Zheng, {\it {Discussions on the
  line-shape of the $X$(4660) resonance}},  {\em Phys. Rev. D} {\bf 100}
  (2019), no.~5 054040, [\href{https://arxiv.org/abs/1906.00356}{{\tt
  arXiv:1906.00356}}].

\bibitem{Wang:2020prx}
J.-Z. Wang, R.-Q. Qian, X.~Liu, and T.~Matsuki, {\it {Are the $Y$ states around
  4.6 GeV from $e^+e^-$ annihilation higher charmonia?}},  {\em Phys. Rev. D}
  {\bf 101} (2020), no.~3 034001, [\href{https://arxiv.org/abs/2001.00175}{{\tt
  arXiv:2001.00175}}].

\bibitem{Ghalenovi:2020zen}
Z.~Ghalenovi and M.~M. Sorkhi, {\it {Spectroscopy of hidden-charm tetraquarks
  in diquark model}},  {\em Eur. Phys. J. Plus} {\bf 135} (2020), no.~5 399.

\bibitem{Karliner:2016zzc}
M.~Karliner, S.~Nussinov, and J.~L. Rosner, {\it {$Q Q \bar Q \bar Q$ states:
  masses, production, and decays}},  {\em Phys. Rev. D} {\bf 95} (2017), no.~3
  034011, [\href{https://arxiv.org/abs/1611.00348}{{\tt arXiv:1611.00348}}].

\bibitem{Chao:1980dv}
K.-T. Chao, {\it {The (cc) - ($\bar{cc}$) (Diquark - Anti-Diquark) States in
  $e^+ e^-$ Annihilation}},  {\em Z. Phys. C} {\bf 7} (1981) 317.

\bibitem{Lloyd:2003yc}
R.~J. Lloyd and J.~P. Vary, {\it {All charm tetraquarks}},  {\em Phys. Rev. D}
  {\bf 70} (2004) 014009, [\href{https://arxiv.org/abs/hep-ph/0311179}{{\tt
  hep-ph/0311179}}].

\bibitem{Chen:2016jxd}
W.~Chen, H.-X. Chen, X.~Liu, T.~Steele, and S.-L. Zhu, {\it {Hunting for exotic
  doubly hidden-charm/bottom tetraquark states}},  {\em Phys. Lett. B} {\bf
  773} (2017) 247--251, [\href{https://arxiv.org/abs/1605.01647}{{\tt
  arXiv:1605.01647}}].

\bibitem{Wu:2016vtq}
J.~Wu, Y.-R. Liu, K.~Chen, X.~Liu, and S.-L. Zhu, {\it {Heavy-flavored
  tetraquark states with the $QQ\bar{Q}\bar{Q}$ configuration}},  {\em Phys.
  Rev. D} {\bf 97} (2018), no.~9 094015,
  [\href{https://arxiv.org/abs/1605.01134}{{\tt arXiv:1605.01134}}].

\bibitem{Debastiani:2017msn}
V.~Debastiani and F.~Navarra, {\it {A non-relativistic model for the
  $[cc][\bar{c}\bar{c}]$ tetraquark}},  {\em Chin. Phys. C} {\bf 43} (2019),
  no.~1 013105, [\href{https://arxiv.org/abs/1706.07553}{{\tt
  arXiv:1706.07553}}].

\bibitem{Wang:2017jtz}
Z.-G. Wang, {\it {Analysis of the $QQ\bar{Q}\bar{Q}$ tetraquark states with QCD
  sum rules}},  {\em Eur. Phys. J. C} {\bf 77} (2017), no.~7 432,
  [\href{https://arxiv.org/abs/1701.04285}{{\tt arXiv:1701.04285}}].

\bibitem{Liu:2019zuc}
M.-S. Liu, Q.-F. Lü, X.-H. Zhong, and Q.~Zhao, {\it {All-heavy tetraquarks}},
  {\em Phys. Rev. D} {\bf 100} (2019), no.~1 016006,
  [\href{https://arxiv.org/abs/1901.02564}{{\tt arXiv:1901.02564}}].

\bibitem{Chen:2020xwe}
H.-X. Chen, W.~Chen, X.~Liu, and S.-L. Zhu, {\it {Strong decays of fully-charm
  tetraquarks into di-charmonia}},
  \href{https://arxiv.org/abs/2006.16027}{{\tt arXiv:2006.16027}}.

\bibitem{Giron:2020wpx}
J.~F. Giron and R.~F. Lebed, {\it {The Simple Spectrum of $c\bar c c\bar c$
  States in the Dynamical Diquark Model}},
  \href{https://arxiv.org/abs/2008.01631}{{\tt arXiv:2008.01631}}.

\bibitem{Karliner:2019vhw}
M.~Karliner and J.~L. Rosner, {\it {Mass inequalities for baryons with heavy
  quarks}},  {\em Phys. Rev. D} {\bf 101} (2020), no.~3 036015,
  [\href{https://arxiv.org/abs/1912.03204}{{\tt arXiv:1912.03204}}].

\bibitem{Karliner:2019hbm}
M.~Karliner and J.~L. Rosner, {\it {LHCb gets closer to discovering the second
  doubly charmed baryon}},  {\em Sci. China Phys. Mech. Astron.} {\bf 63}
  (2020), no.~2 221064, [\href{https://arxiv.org/abs/1912.01963}{{\tt
  arXiv:1912.01963}}].

\bibitem{Ablikim:2019hff}
{\bf BESIII} Collaboration, M.~Ablikim et~al., {\it {Future Physics Programme
  of BESIII}},  {\em Chin. Phys. C} {\bf 44} (2020), no.~4 040001,
  [\href{https://arxiv.org/abs/1912.05983}{{\tt arXiv:1912.05983}}].

\bibitem{Ablikim:2017lct}
{\bf BESIII} Collaboration, M.~Ablikim et~al., {\it {Precision measurement of
  the $e^{+}e^{-}~\rightarrow~\Lambda_{c}^{+} \bar{\Lambda}_{c}^{-}$ cross
  section near threshold}},  {\em Phys. Rev. Lett.} {\bf 120} (2018), no.~13
  132001, [\href{https://arxiv.org/abs/1710.00150}{{\tt arXiv:1710.00150}}].

\bibitem{Wang:2020gmd}
X.-Y. Wang, Q.-Y. Lin, H.~Xu, Y.-P. Xie, Y.~Huang, and X.~Chen, {\it {Discovery
  potential for the LHCb fully-charm tetraquark $X(6900)$ state via $\bar{p}p$
  annihilation reaction}},  \href{https://arxiv.org/abs/2007.09697}{{\tt
  arXiv:2007.09697}}.

\bibitem{Wang:2020wrp}
J.-Z. Wang, D.-Y. Chen, X.~Liu, and T.~Matsuki, {\it {Producing fully-charm
  structures in the $J/\psi$-pair invariant mass spectrum}},
  \href{https://arxiv.org/abs/2008.07430}{{\tt arXiv:2008.07430}}.

\bibitem{Azizi:2019xla}
K.~Azizi, S.~Agaev, and H.~Sundu, {\it {The Scalar Hexaquark $uuddss$: a
  Candidate to Dark Matter?}},  {\em J. Phys. G} {\bf 47} (2020), no.~9 095001,
  [\href{https://arxiv.org/abs/1904.09913}{{\tt arXiv:1904.09913}}].

\bibitem{Farrar:2020zeo}
G.~R. Farrar, Z.~Wang, and X.~Xu, {\it {Dark Matter Particle in QCD}},
  \href{https://arxiv.org/abs/2007.10378}{{\tt arXiv:2007.10378}}.

\end{thebibliography}\endgroup

\end{document}